\documentclass[twocolumn]{aastex63}

\newcommand{\teff}{T_{\rm eff}}
\newcommand{\kms}{\hbox{km\,s$^{-1}$}}
\newcommand{\hei}{He\,{\sc i}}
\newcommand{\heii}{He\,{\sc ii}}
\newcommand{\hi}{H\,{\sc i}}

\newcommand{\hb}{H$\beta$}
\newcommand{\hg}{H$\gamma$}
\newcommand{\hd}{H$\delta$}
\newcommand{\vsini}{$v\,$sin$\,i$}
\newcommand{\azv}{AzV~493}
\newcommand{\msun}{$\rm M_\odot$}

\usepackage{mathrsfs}
\usepackage{graphics}

\newcommand{\new}[1]{\textcolor{black}{#1}}
\accepted{January 23, 2023; to appear in the Astrophysical Journal}
\begin{document}

\title{Strong Variability in AzV 493, an Extreme Oe-Type Star in the SMC}

\author[0000-0002-5808-1320]{M. S. Oey}
\affiliation{Astronomy Department, University of Michigan, 1085 South University Ave., Ann Arbor, MI, 48109, USA}
% \author{Norberto Castro\altaffilmark{2}}
\author[0000-0003-0521-473X]{N. Castro}
\affiliation{Leibniz-Institut für Astrophysik Potsdam (AIP), An der Sternwarte 16, 14482, Potsdam, Germany}
\author[0000-0002-6718-9472]{M. Renzo}
\affiliation{Center for Computational Astrophysics, Flatiron Institute, 162 5th Ave, New York, NY 10010, USA}
\author[0000-0001-7046-6517]{I. Vargas-Salazar}
\affiliation{Astronomy Department, University of Michigan, 1085 South University Ave., Ann Arbor, MI, 48109, USA}
\author[0000-0003-0696-2983]{M. W. Suffak}
\affiliation{Department of Physics and Astronomy, Western University, London, ON N6A 3K7, Canada}
\author[0000-0002-3218-2684]{M. Ratajczak}
\affiliation{Astronomical Observatory, University of Warsaw, Al. Ujazdowskie 4, 00-478 Warszawa, Poland}
\author[0000-0002-3380-3307]{J. D. Monnier}
\affiliation{Astronomy Department, University of Michigan, 1085 South University Ave., Ann Arbor, MI, 48109, USA}
\author[0000-0002-0548-8995]{M. K. Szymanski}
\affiliation{Astronomical Observatory, University of Warsaw, Al. Ujazdowskie 4, 00-478 Warszawa, Poland}
\author{G. D. Phillips}
\affiliation{Astronomy Department, University of Michigan, 1085 South University Ave., Ann Arbor, MI, 48109, USA}
\author[0000-0002-3950-5386]{N. Calvet}
\affiliation{Astronomy Department, University of Michigan, 1085 South University Ave., Ann Arbor, MI, 48109, USA}
\author[0000-0002-7155-679X]{A. Chiti}
\affiliation{Department of Astronomy \& Astrophysics, University of Chicago, 5640 S Ellis Avenue, Chicago, IL 60637, USA}
\affiliation{Kavli Institute for Cosmological Physics, University of Chicago, Chicago, IL 60637, USA}
\author[0000-0003-4376-2841]{G. Micheva}
\affiliation{Astronomy Department, University of Michigan, 1085 South University Ave., Ann Arbor, MI, 48109, USA}
\affiliation{Present address:  Leibniz-Institut für Astrophysik Potsdam (AIP), An der Sternwarte 16, 14482, Potsdam, Germany}
\author[0000-0002-0470-0800]{K. C. Rasmussen}
\affiliation{Astronomy Department, University of Michigan, 1085 South University Ave., Ann Arbor, MI, 48109, USA}
\affiliation{Present address:  Astronomy Department, University of Washington, Box 351580, Seattle, WA  98195, USA}
\author[0000-0002-2522-8605]{R. H. D. Townsend}
\affiliation{Astronomy Department, University of Wisconsin, Madison, WI 53706, USA}

\begin{abstract}
We present 18 years of OGLE photometry together with spectra obtained over 12 years, revealing that the early Oe star \azv\ shows strong photometric ($\Delta I < 1.2$ mag) and spectroscopic variability 
with a dominant, 14.6-year pattern and $\sim$40-day 
{oscillations.  We estimate
stellar parameters} $T_{\rm eff} = 42000$ K, $\log L/L_\odot=5.83 \pm
0.15$, $M/M_\odot = 50\pm 9$, and \vsini\ $=370\pm 40$ \kms.  Direct
spectroscopic evidence shows episodes of both gas ejection and infall.
There is no X-ray detection, and it is likely a runaway star.
\azv\ may {have} an unseen companion on a highly eccentric
{($e>0.93$) orbit}.
We propose that
close interaction at periastron excites ejection of the decretion
disk, whose variable emission-line spectrum
{suggests separate inner and outer components, with an optically
thick outer component} obscuring both the stellar
photosphere and the emission-line spectrum of the inner disk
at early phases in the
{
photometric cycle.  It is plausible that \azv's}
mass and rotation have been enhanced by
binary interaction followed by the core-collapse supernova explosion
of the companion, 
{which now could be}
either a black hole or neutron star.
This system in the Small Magellanic Cloud can
potentially shed light on
{OBe decretion disk formation and evolution, 
massive binary evolution, and compact binary
progenitors.}
\end{abstract}



\section{Introduction} \label{sec:intro}

Binary interactions are now understood to be a fundamental component of massive star evolution, and they are the progenitors of a wide variety of energetic phenomena including high-mass X-ray binaries (HMXBs), 
ultra-luminous X-ray sources (ULXs), stripped-envelope core-collapse supernovae (SNe), and gravitational wave events.
{A consensus is emerging that classical OBe stars appear to
  originate from close massive binary systems, wherein they} have spun up through mass
and angular momentum transfer from their {mass donors} (e.g,
\citealt{pols:91, vinciguerra:20, bodensteiner:20}, see also
\citealt{2013A&ARv..21...69R} for a review).
When donor stars 
subsequently explode as supernovae, resulting post-explosion bound binaries are more likely to be eccentric, since they result from tight binaries \citep[e.g.,][]{brandt:95, tauris:98, renzo:19walk}.
{Thus, } a substantial subset of classical OBe stars are likely to have 
eccentric orbits.
In this paper, we present photometric and spectrocopic time-series
data showing that the star \azv\ exhibits dramatic variability and
{may be an}
eccentric binary system.

AzV 493 \citep{Azzopardi1975} or [M2002]SMC-77616
\citep{2002ApJS..141...81M} was identified as an extreme, 
classical Oe star by \cite{2016ApJ...819...55G}.  In that
work, it was found to be the earliest classical Oe star in our sample
of field OB stars in the Small Magellanic Cloud (SMC), based on a
spectrum obtained in 2009 that shows double-peaked emission, not only
in the Balmer lines, but also in \hei\ and \heii\,$\lambda4686$, the latter
feature being rarely observed in other Oe stars \citep{1974ApJ...193..113C}.
Specifically, it is classified as an Ope star, indicating that the
\hei\ absorption lines show infilled emission \citep{Sota2011}.  

{As an extreme object, \azv\ offers unique opportunities to study massive binary
  evolution and decretion disk formation, structure, and dynamics.}
% It may be a heartbeat star, and if so, it would be one of the most extreme such objects known to date.
{Section~\ref{sec:phot0} presents the unusual light curve and periodicity, and Section~\ref{sec:spec} presents our multi-epoch spectroscopy with resulting derived stellar parameters and individual spectral features.  We then present two possible models for the \azv\ system in Sections~\ref{sec:model1} and \ref{sec:diskdiscussion}, one based on ejection of an optically thick disk near periastron; and another based on disk growth and disruption.  Section~\ref{sec:binary} discusses the
likely binary origin of the system, and Section~\ref{sec:sum} summarizes our findings.}

\section{Photometric light curve} \label{sec:phot0}

\subsection{{Long-term light curve}} \label{sec:phot}

\begin{figure*}
    \includegraphics[width=\textwidth]{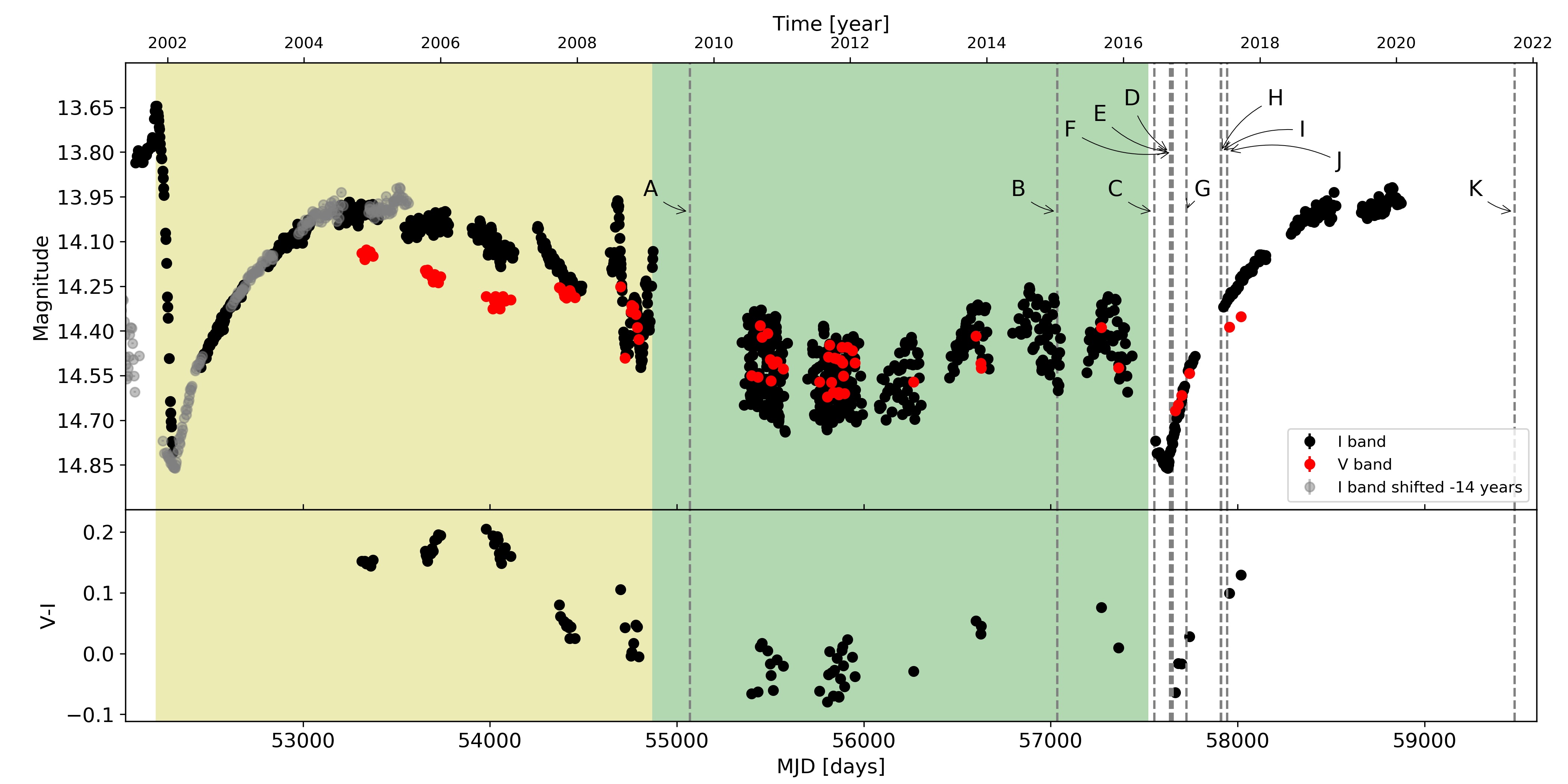}
    \caption{AzV~493 OGLE light curves in $I$ (black) and $V$ (red)
      bands.  The last segment of the $I$-band curve is overplotted
      (light grey dots) on the beginning of the dataset phase 14.6
      years (5311 days) earlier.  $V-I$ is shown in the lower panel.
      The dashed lines mark the epochs for the observed spectra, assigned alphabetically in chronological sequence.  The
      green shaded regions show consecutive 2656-day segments starting
      with the light curve maximum in 2001.  
	\label{fig:ogle_LC}}
\end{figure*}

The $I$ and $V$-band light curves of AzV~493 from the OGLE Project \citep{2008AcA....58...69U, Udalski2015} are presented
in Figure~\ref{fig:ogle_LC}.  The $I$-band shows a short eruption with the peak of the light curve on { MJD 52212,} followed by an abrupt decline of approximately 1.2 mag, to a minimum on MJD 52303 in early 2002.
After this, the star eventually recovers its original luminosity.
Another photometric minimum is seen in 2016 on MJD 57626, followed by the same brightening pattern.
The gray symbols in Figure~\ref{fig:ogle_LC} show the $I$-band photometry from the 2016 cycle overplotted on the data from 2002 cycle.  This shows that the minimum luminosity and subsequent increase are quantitatively identical, although the photometry immediately preceding the minimum differs.  
Cross-correlating these segments yields a long-cycle period of 5311 days (14.55 years).
There is no evidence of a similar eruption preceding the minimum in the 2016 cycle on the same 91-day timescale, although the photometry is incomplete in this range.  

% First minimum: MJD 52303.0712799998    I-mag= 14.808
% Second :      MJD 57626.331809999887  I-mag= 14.8610

After the minimum, the brightness increases and then starts to gradually decrease again, over a
period of several years.
Approximately in 2008, AzV~493 appears to go into a multiple outburst event.
After this, the light curve drastically changes, showing a multi-mode pulsation
behavior that evolves with time  (Section~\ref{Sect:Pulsation}).
The pulsation ends with another 0.2 -- 0.3 mag drop, followed by a
steady increase, repeating the light curve
cycle that started in 2002, 14.55 years before.

\begin{figure*}
% \vspace*{-0.7in}
\includegraphics[width=\textwidth]{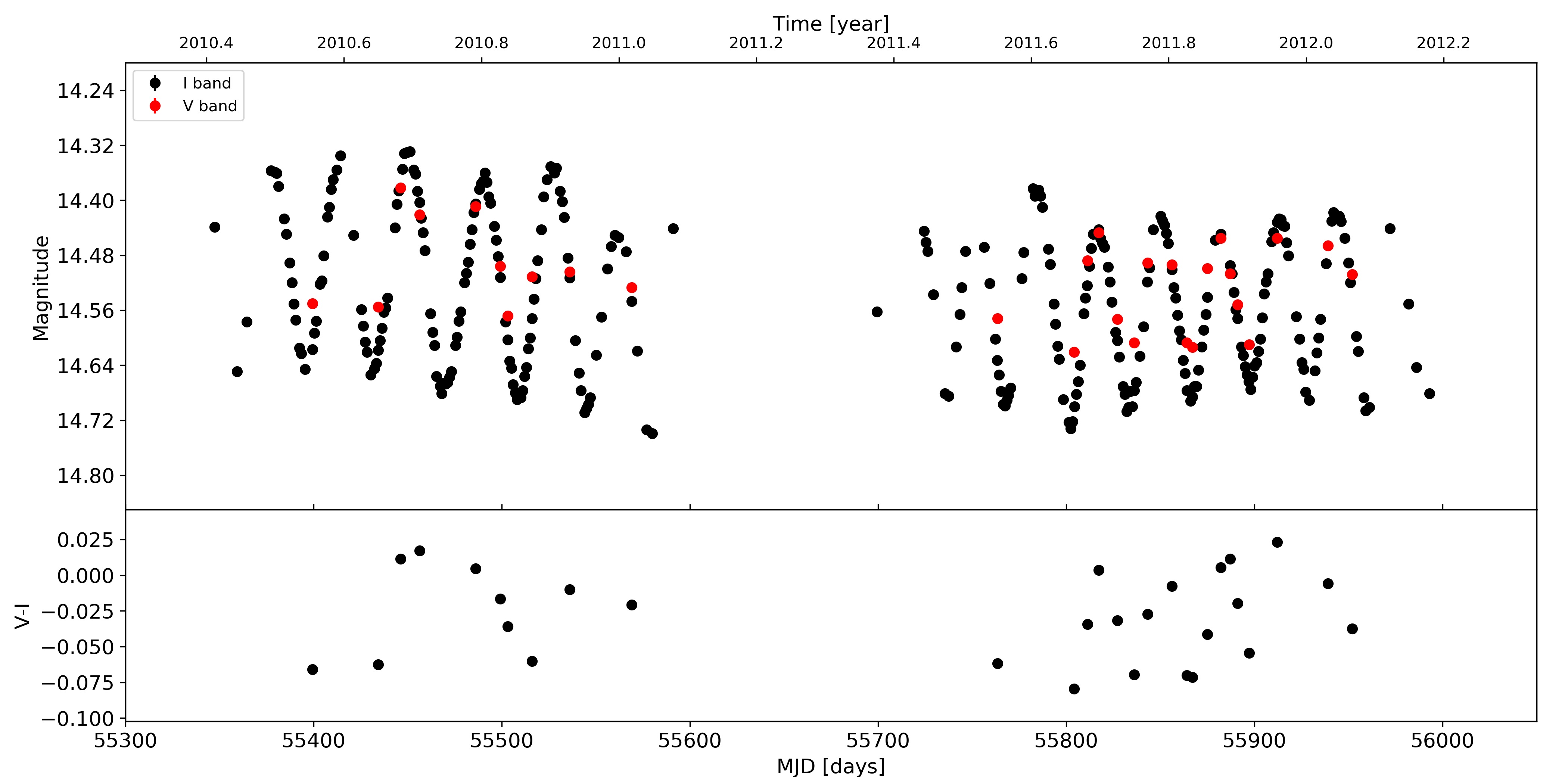}
\vspace*{-0.7in}
	\caption{Zoom on light curve (top) showing $\sim 40$-day {
            oscillations}, and color variation (bottom).  \label{fig:pulseloop}}
\end{figure*}

\begin{figure}[h]
	\includegraphics[width=3in]{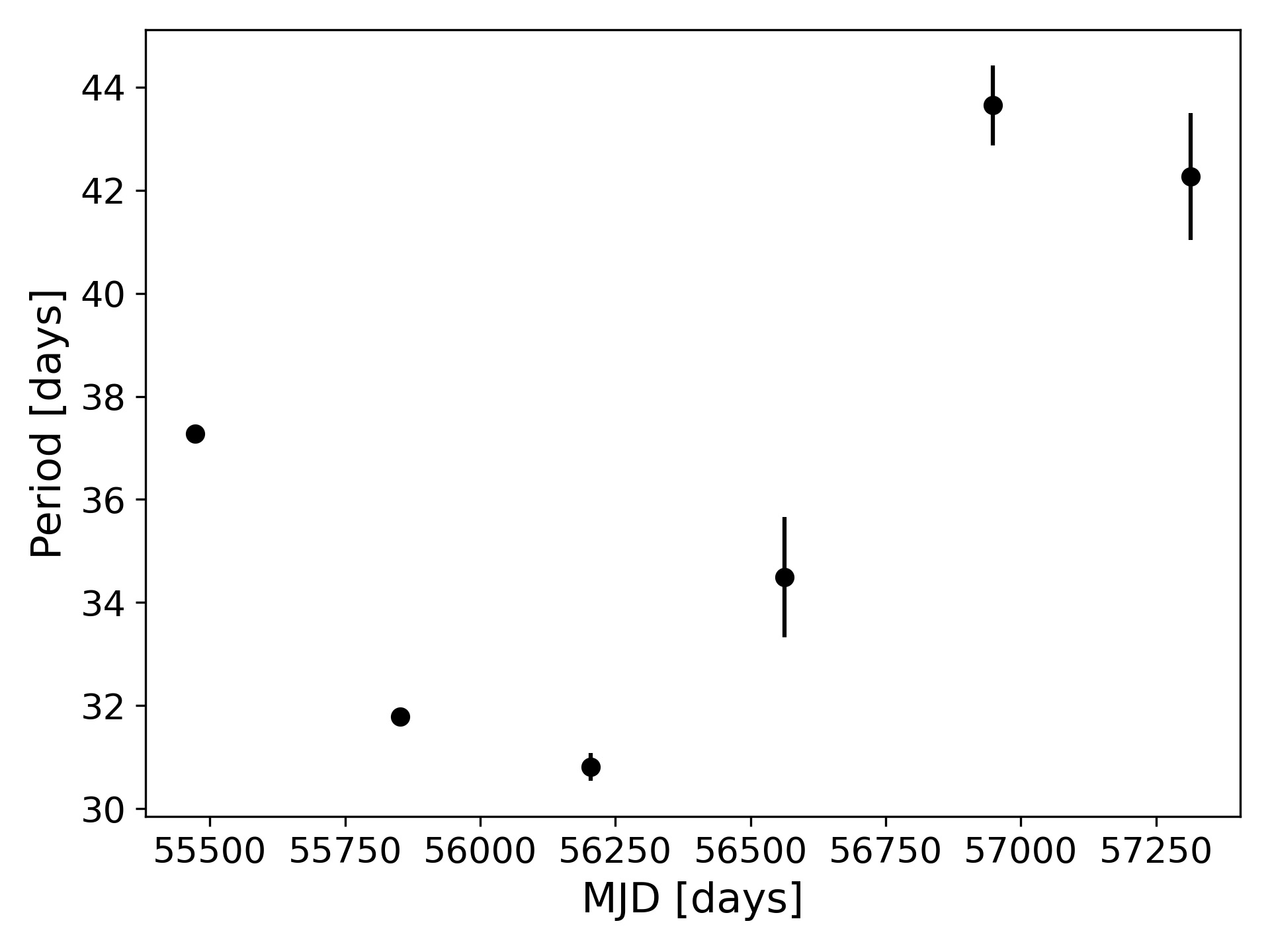}
%  \vspace*{-1in}
	\caption{Fitted periods for the six contiguous OGLE datasets between $\sim$ 2010 -- 2016, as a function of time.
        }
	\label{fig:GLSperiods}
\end{figure}

\begin{figure*}
	\includegraphics[width=2.9in]{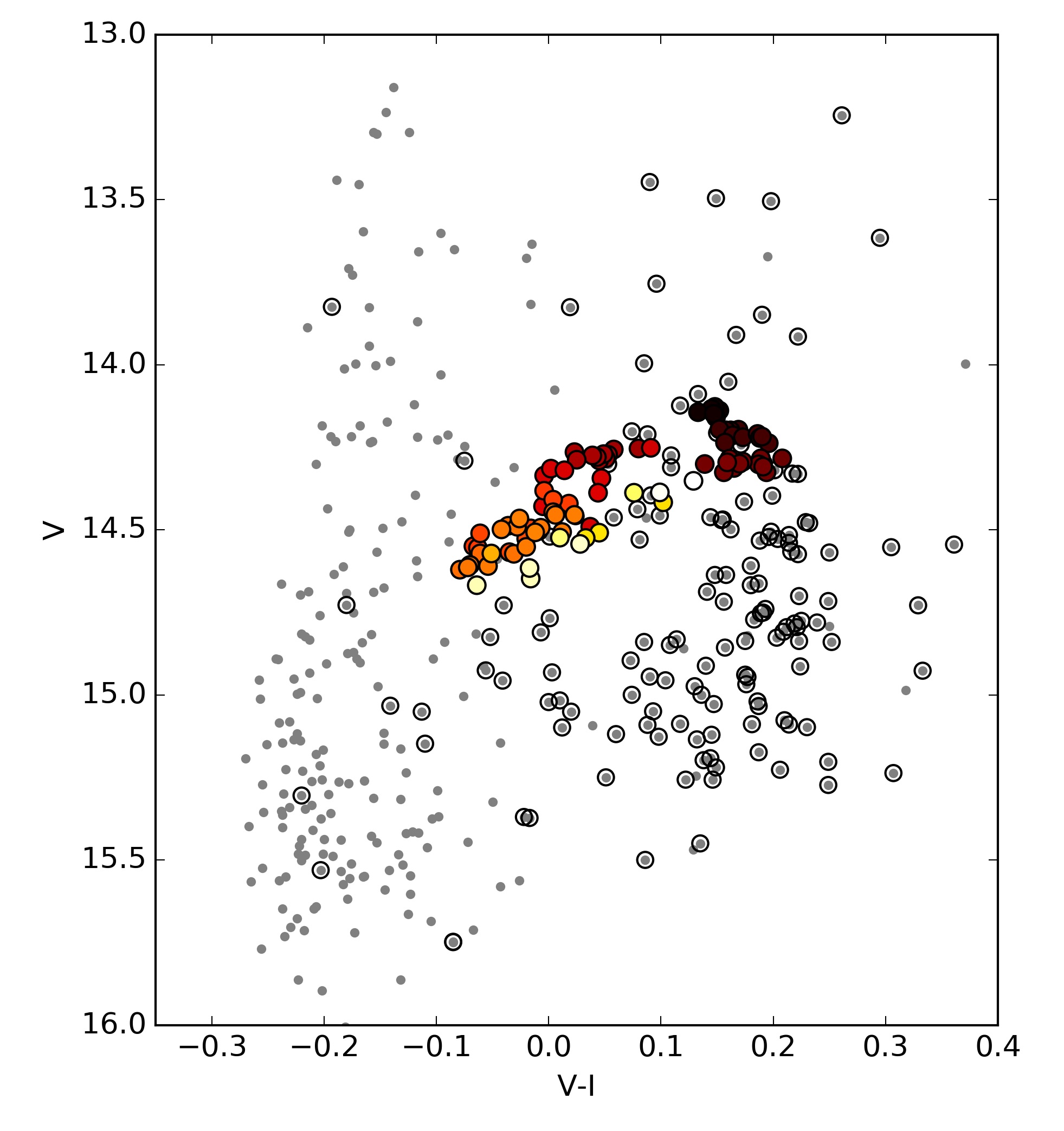}
	\includegraphics[width=4.1in]{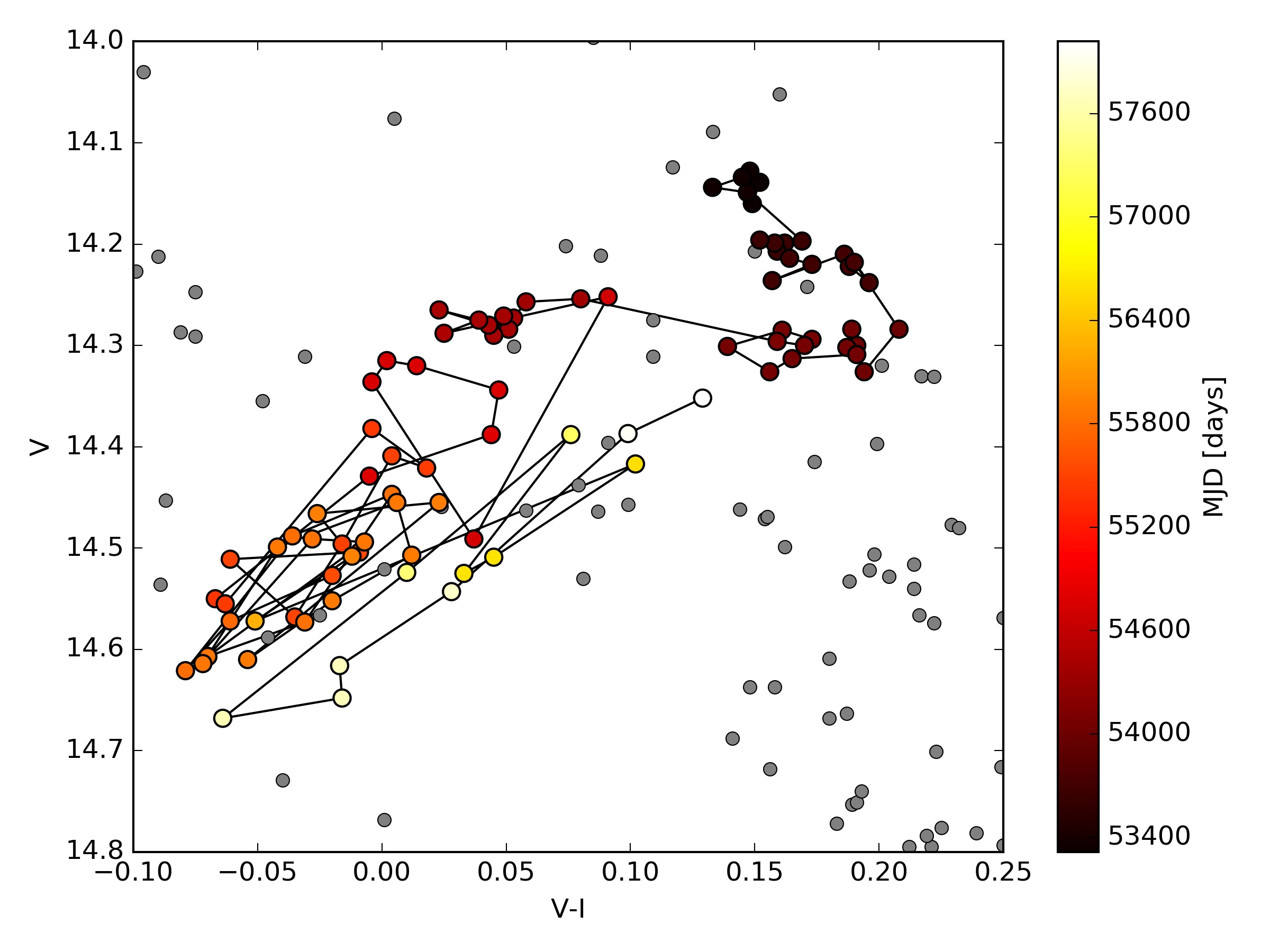}
	\caption{Color-magnitude diagram (CMD) based on available $V$- and
          $I$-band OGLE photometry (see Figure~\ref{fig:ogle_LC}).  The variation of AzV~493 in the
          CMD is colored according to the MJD, and compared to
          single-epoch OGLE photometry \citep{2012AcA....62....1P} for
          the RIOTS4 OB-star sample \citep{2016ApJ...817..113L} (grey
          dots).  Objects classified as OBe by
          \citet{2016ApJ...817..113L} are highlighted with circles.
          The right panel is a zoom of the same
          data around the track of \azv.
          \label{fig:ogle_CMD}
        }
\end{figure*}

\subsection{Photometric Oscillations}
\label{Sect:Pulsation}

Figure~\ref{fig:pulseloop} shows short-term variability on the order of 30 -- 45 days.  We quantify the evolution 
of these oscillations
seen in the $I$-band light curve using Generalized Lomb-Scargle periodograms
\citep{Zechmeister2009}
for the six contiguous OGLE datasets from 2010 -- 2016 (Figure~\ref{fig:ogle_LC}).
The individual fits to these six ranges are shown in Appendix~A.  Comparison of the periods shown in Figure~\ref{fig:GLSperiods} with the light curve (Figure~\ref{fig:ogle_LC}) shows that they qualitatively appear to correlate with stellar brightness.  

The OGLE survey provides $V$-band magnitudes for a subset of the survey epochs, which are shown in red in Figure~\ref{fig:ogle_LC}.
Figure~\ref{fig:ogle_CMD} displays the color-magnitude diagram (CMD) in $V$ vs $V-I$
for those days where both bands were observed.
Figure~\ref{fig:ogle_CMD}a compares AzV~493's color variations with
data for the remainder of the RIOTS4 sample stars \citep{2016ApJ...817..113L}.  The latter correspond to
single-epoch photometry  from the OGLE catalog of \cite{2012AcA....62....1P}.  
Those stars classified as OBe stars by  \citet{2016ApJ...817..113L} are marked in the plot.
The blue plume of non-emission-line stars is clearly separated from the cloud of OBe stars at redder colors in the CMD, a
phenomenon already known from different photometric bands \citep[e.g.,][]{2010AJ....140..416B, Castro2018}.
The color variation of AzV~493 spans almost the entire range of $V-I$ colors covered by  the emission-line stars.

Figure~\ref{fig:ogle_CMD}b shows a zoom in the CMD with the path of AzV~493 traced out.
The star appears red during the broad peak of the light curve around 2006 (Figure~\ref{fig:ogle_LC}), and then moves
to bluer colors reaching the bluest $V-I$ color during the pulsation
phase.  Approximately in 2017, when the light curve is
brightening after the minimum, AzV~493 shows redder colors again, moving to the
original position observed in 2005 with $V-I\sim 0.18$.

{Similar, semi-periodic variability with timescales on the order of
weeks to months is seen in many other OBe stars, and their origin is
unknown \citep[e.g.,][]{Labadie2017}.
Proposed explanations include forms of non-radial pulsations of the
star and transitory or orbiting density enhancements in the disk,
which may be the most likely scenario.
The associated cyclical}
variation in the CMD (Figure~\ref{fig:ogle_CMD})
is also consistent with some kind of stellar radial pulsation.
{This is supported by the correlation between period and
luminosity (cf. Figures~\ref{fig:GLSperiods} and \ref{fig:ogle_CMD}).
In that case,}
the relatively long period implies that they
{could be an induced gravity mode} or pulsational instability.
However,
{there are many other possible explanations, perhaps including
  interactions with another star in a close orbit.}
We note that
\citet[][see also \citealt{2013A&ARv..21...69R}]{2006A&A...456.1027D} reported similar loop-like excursions in
the CMD of other OBe stars, and ascribed the anti-clockwise variation
to the formation and dissipation of the circumstellar decretion disks
in those objects. 

\subsection{{Light curve period}}\label{sec:period}

{It is possible that} the multiple-outburst event in 2008 -- 2009 {may represent} another periastron.
Figure~\ref{fig:ogle_LC} shows the 5311-day cycle initiated at the
light-curve peak at {MJD 52212} instead of at the minima.  We see that
the mid-cycle occurs during this multiple-outburst event, although due
to the OGLE observing cadence, it is unclear whether it occurs near
the end or near the middle.  In Section ~\ref{sec:spec} below, we show
that the spectrum obtained around this time, Epoch~A
(Figure~\ref{fig:ogle_LC}), shows an unusually strong emission-line
spectrum, consistent with {maximum} disk activation and flaring.
However, the light curve does not repeat the cycle {minimum} seen
in 2002 and 2016,  and OBe stars are known to show temporary
  outbursts of activity \citep[e.g.,][]{Labadie2017, Baade2018}.  

Thus, it is not clear whether 2008 -- 2009 {corresponds to the
mid-cycle or not.}
The light curve does not repeat regularly in detail, and we caution
that the period, if the system is a binary, is uncertain.  Assuming
that { there is indeed a fundamental physical period},
the {same phases} may not all generate the same observational
signatures, which may depend on {other factors such as} disk orientation and/or varying physical processes.
{In what follows, we adopt a system period of 5311 (2656) days, or
  14.55 (7.28) years, where the values in parentheses allow for the
  possibility that the period may be half of the long cycle.
}

\section{Spectroscopy} \label{sec:spec}

\noindent
\begin{table*}
	\caption{Spectroscopic Observations of \azv}
	\label{tab:obs_log}
	\begin{center}
		\begin{footnotesize}
			\begin{tabular}{cccrrccccl}
				\hline
			\hline
		Epoch & Date [UTC] &MJD  & S/N & $R$~~~ & Wavelength & Phase\tablenotemark{a} & RV &
		$\Delta v(\rm H\beta) $\tablenotemark{b}  & Instrument  \\
		 &  & &  &  & Range [\AA] & & ($\kms$) & (km\,s$^{-1}$) &   \\
	\hline
A & 2009-08-26T01:43:36.0 & 55069.071944 & 140 & 3000 &3825--5422 &  0.538 (0.076)  & $152\pm 200$ & 279 & IMACS \\
B & 2015-01-14T02:12:03.0 & 57036.091701 & 120 & 28000 &3362--9397  &   0.908 (0.817) & $192\pm 18$ & (213)\tablenotemark{c} &  MIKE \\
C & 2016-06-15T07:47:54.3 & 57554.324935 & 130 & 3000 &  3879--5479 &  0.006 (0.012) & $171\pm 60$ & 346 &  IMACS \\
D & 2016-09-08T01:42:08.0 & 57639.070926 & 60 & 28000&  4079--4466 &  0.022 (0.044) & $217\pm 50$  & \nodata &  M2FS \\
E & 2016-09-11T02:49:33.0 & 57642.117743 & 90 & 28000 &4080--4465 &  0.022 (0.045) & $239\pm 46$ & \nodata &  M2FS \\
F & 2016-09-22T05:36:51.0 & 57653.233924 & 150 & 28000 & 3538--9397 &  0.024 (0.049) & $192\pm 29$ & 334 & MIKE \\
G & 2016-12-04T04:09:41.5 & 57726.173397 & 110 & 3000 &  3862--5458 &  0.038 (0.076) & $243\pm 38$ & 319 &IMACS \\
H & 2017-06-05T06:35:11.2 & 57909.274435 & 50 & 3000 & 3871--5471 &  0.073 (0.145) & $235\pm 54$ & 322 &IMACS \\
I & 2017-06-07T08:08:18.9 & 57911.339108 & 130 & 1300 & 3900--8000 &  0.073 (0.146) & $231\pm 83$ & 295 & IMACS\tablenotemark{d} \\
J & 2017-07-10T09:05:00.5 & 57944.378478 & 190 &3000 & 3854--5468 &  0.079 (0.159)& $181\pm 39$  & 303 &  IMACS \\
K & 2021-09-25T07:38:18.0 & 59482.318264 & 210 & 28000 & 3362--9397 &  0.369 (0.738) & $183\pm 17$  & 289 &  MIKE \\
\hline
\end{tabular}
\end{footnotesize}
\end{center}
\tablenotetext{a}{Phase relative to the light curve peak at {MJD
  52212 (54868), adopting a period of 5311 (2655.5) days.} }
\tablenotetext{b}{H$\beta$ peak separation obtained by fitting two gaussians with fixed width of 2 \AA\ ($\sim 120\ \kms$). }
\tablenotetext{c}{Epoch B does not show a double-peaked profile (see Figure~\ref{fig:Hbpeak} and Section~\ref{sec:epochBF}); 
the value for $\Delta v(\rm H\beta)$ assumes that two components exist, as they do for other epochs. }
\tablenotemark{d}{Epoch I was observed with the f/2 camera while the other IMACS observations were obtained with the f/4 camera.}
\end{table*}

\begin{figure*}
%	\rotatefig{90}{Spectra_V250522_b}{8.5in}{}
{\includegraphics[angle=90,height=\textheight]{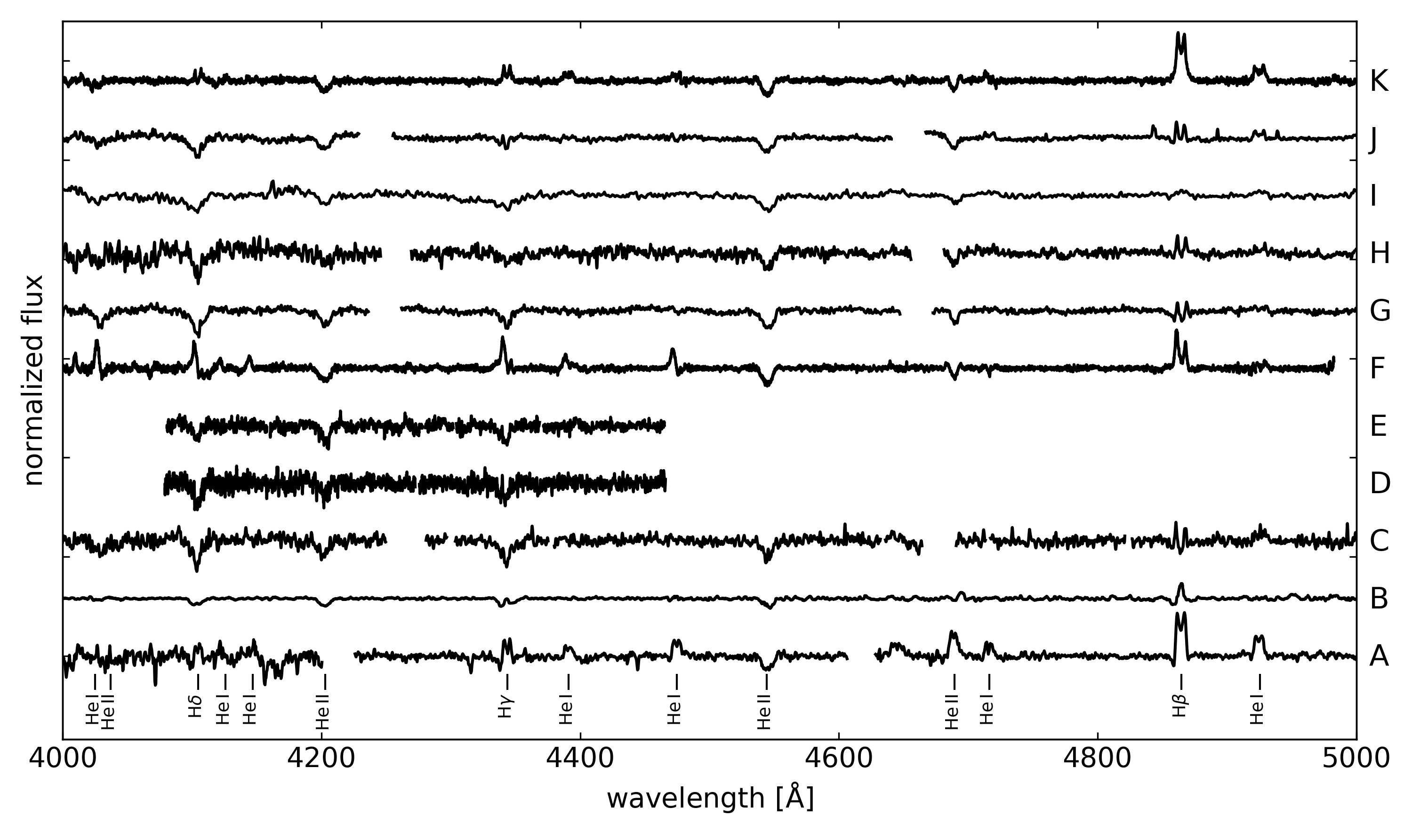}}
\caption{AzV~493 multi-epoch spectroscopic observations sorted by MJD
{and normalized to the continuum.}  Epoch~I is low resolution
    (Table~\ref{tab:obs_log}).  
	\label{fig:all}}
\end{figure*}

Spectroscopic observations of \azv\ were obtained in the course of the RIOTS4 spectroscopic survey of field OB stars in the SMC \citep{2016ApJ...817..113L}, and follow-up radial velocity monitoring of the SMC Wing region (Vargas-Salazar et al. 2023, in preparation).  The
observations were carried out using the Magellan
telescopes at Las Campanas, Chile.  Three different spectrographs were used:
IMACS \citep{2003SPIE.4841.1727B}, MIKE \citep{2003SPIE.4841.1694B}
and M2FS \citep{2012SPIE.8446E..4YM}.  Table~\ref{tab:obs_log} gives
details of our spectroscopic observations, including
the modified Julian day (MJD), signal-to-noise, spectral resolution,
spectral range, phase in the light curve cycle, radial velocity,
\hb\ peak separation (Section~\ref{sec:VR}), and instrument used.
Figure~\ref{fig:all} 
displays the 11 spectra in chronological sequence, labeled A -- K as shown.

IMACS was operated by default in multi-slit mode with the  f/4 camera and 1200
lines/mm grating,
which provides a resolving power of $R\,\sim3000$ and a
wavelength coverage spanning $\sim$3800 -- 5200\,\AA.  
One observation (Epoch~I) was observed with the f/2 camera, resulting in lower resolution (Table~\ref{tab:obs_log}).
The reduction was performed using the {\sc
  cosmos} pipeline\footnote{http://code.obs.carnegiescience.edu/cosmos.}.
MIKE data were obtained using a 1$\arcsec$
slit width for a spectral resolution of $R\,\sim28000$,
covering the wavelength range $\sim$3600 -- 10000\,\AA.
The spectra were processed with the
the Carnegie Python ({\sc CarPy}\footnote{http://code.obs.carnegiescience.edu/mike}) pipeline software
\citep{2000ApJ...531..159K,2003PASP..115..688K}, except for Epoch~B, which was extracted using 
{\sc IRAF}\footnote{{\sc IRAF} was distributed by the National Optical Astronomy Observatory, which was managed by the Association of Universities for Research in Astronomy (AURA) under a cooperative agreement with the National Science Foundation.}.
M2FS data were observed using a custom filter yielding
$\sim$4080 -- 4470 \AA\ wavelength coverage at $R\,\sim28000$.  The data were
processed following the standard steps in fiber spectroscopic reduction
using  IRAF/PyRAF tasks implemented within python and  designed for this
instrument (see \citealt{2015ApJ...808..108W}).

{Figure~\ref{fig:all} shows strong variability in the spectrum of \azv.}
The weaker epochs show a typical OBe spectrum, with only H$\beta$
showing double-peaked emission, and \hg\ and H$\delta$ absorption
features showing evidence of infill; {whereas} Epochs~A, B, and K
show stronger emission-line spectra, {with \hg\ and \hei\ often in emission}.  Epoch~F shows
strong, {high-order Balmer} emission and inverse P-Cygni features.  These epochs will
be discussed in Sections~\ref{sec:epochA} -- \ref{sec:epochBF}. 

\subsection{Stellar fundamental parameters}\label{sec:stellparams}

The photospheric \heii\ lines $\lambda$4200, $\lambda$4541, and $\lambda$5411 lines at all epochs
confirm the early O spectral type assigned by
\cite{2016ApJ...819...55G}.
To improve S/N in the \heii\,$\lambda4541$ absorption line, we combine epochs {C}, {G}, {H} and {J},
which are all IMACS spectra obtained in 2016 -- 2017.
We use this composite spectrum to estimate the projected rotational
velocity  (\ensuremath{{\upsilon}\sin i}) using  the {\sc iacob-broad} code
\citep{2014A&A...562A.135S,2007A&A...468.1063S}.  We obtain
$\upsilon\sin i=370\pm40\,$km\,s$^{-1}$.
As discussed in Section~\ref{sec:model1}, the angle of inclination $i$ is likely high, based on the amount of obscuration from the disk, and so the rotational velocity might be $\lesssim 450\ \kms$.

\begin{figure*}
	%	\figurenum{1}
	\plotone{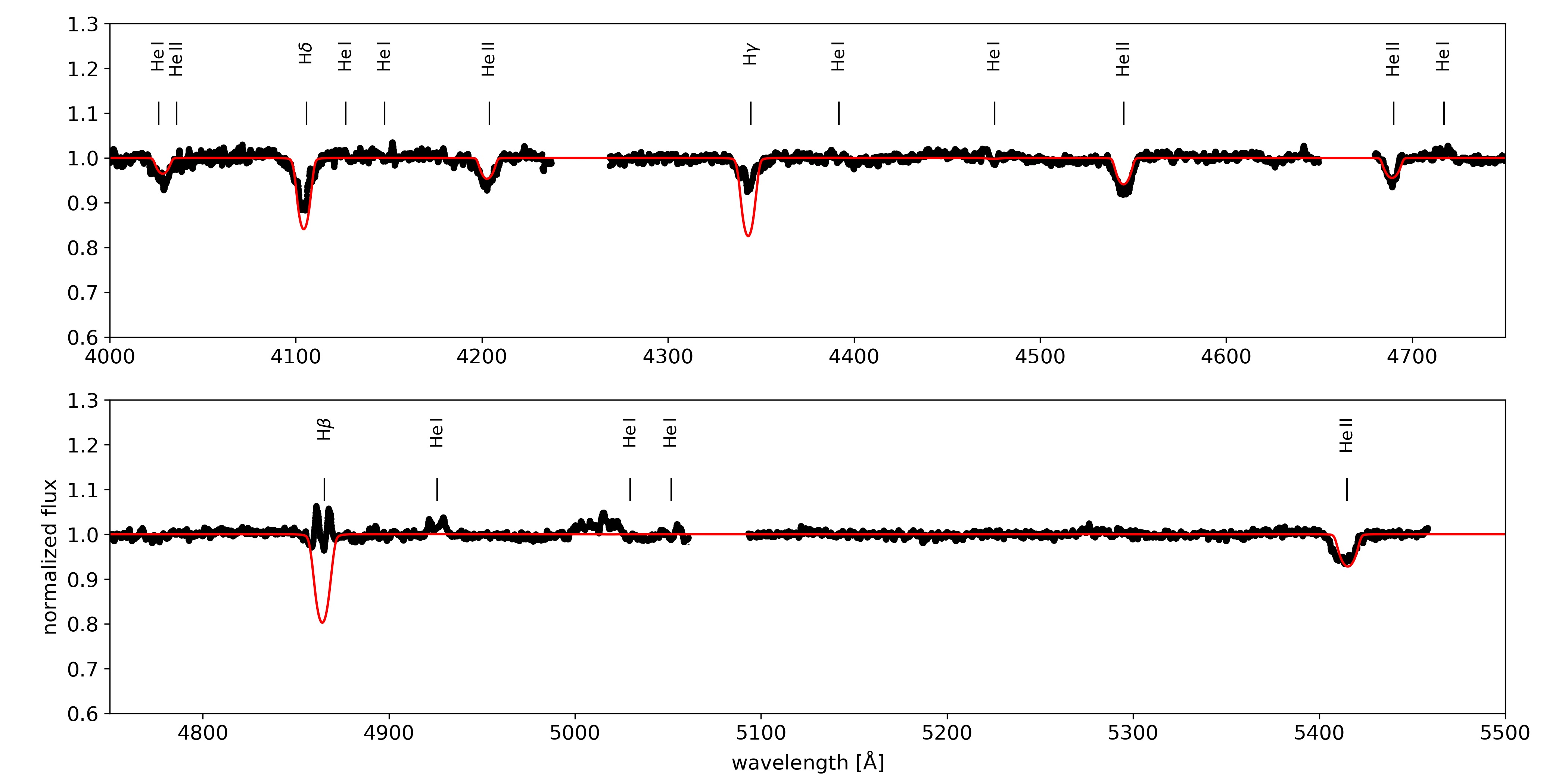}
	\caption{Spectroscopic analysis of the composite IMACS
          spectrum from epochs C, G, H and J  (black;
          cf. Fig.~\ref{fig:all}).  The best {\sc fastwind}
          \citep{1997A&A...323..488S,2005A&A...435..669P,2012A&A...543A..95R}
          stellar atmosphere synthetic model is overplotted (red).
          The main transitions used in the analysis are marked.
          \label{fig:spec}}
\end{figure*}

The combined spectrum was modelled using the stellar atmosphere code {\sc fastwind} \citep{1997A&A...323..488S,2005A&A...435..669P,2012A&A...543A..95R},
using the same technique and stellar grid
described in \cite{Castro2018}.
The cores of the Balmer lines are omitted from the fit to ameliorate contamination from disk emission.
Our best model yields effective temperature $\teff =  42000\,$K and surface gravity $\log g = 3.4$\,dex, which
reproduce the main \hei\ and \heii\  lines (Figure~\ref{fig:spec}).
{Since \hei\ photospheric features are not detected, this $\teff$ may
  be a lower limit.}
The derived temperature is consistent with an {O3-5 spectral type
\citep{Martins2021},}
matching the early O-type classification of AzV 493 \citep{2016ApJ...817..113L}.
However, we caution that the wings of the Balmer lines, which are the main
spectroscopic anchors for deriving the surface gravity, may be affected by the
circumstellar emission, resulting in an underestimate of $\log g$, as found for OBe stars by \citet{Castro2018}.  

The stellar luminosity was calculated using the optical and IR photometry for AzV 493
\citep{2002ApJS..141...81M,2006AJ....131.1163S},
adopting a distance to the SMC of 62.1 kpc \citep{2014ApJ...780...59G}
and the synthetic {\sc fastwind}
spectral energy distribution (SED) derived above. We explored the extinction curves published
by \cite{2007ApJ...663..320F} until the observed photometry was reproduced  by the
{\sc fastwind} synthetic SED. We obtain a luminosity $\log L/L_\odot=5.83\pm0.15$ and radius $R_\star/R_\odot = 15\pm3$, in agreement with
the expected values for an early O-type star of luminosity class III -- V
\citep[e.g.][]{2005A&A...436.1049M}. We compare the position
of AzV 493 in the  Hertzsprung–Russell diagram with the rotating evolutionary tracks by \cite{2011A&A...530A.115B} for
SMC metallicity. Based on the  $\teff$ and $L/L_\odot$ and their respective uncertainties, we estimate that the stellar mass is $M/M_\odot = 50\pm9$.
{If the observed luminosity is overestimated by the inferred $\log
  g$, or includes a
contribution from a non-compact binary companion and/or the disk
continuum, then the stellar mass may be somewhat overestimated; for
reference, a factor of two overestimate in luminosity implies $M/M_\odot\sim 40$.}

\subsection{\hb\ emission-line profile}\label{sec:VR}

\begin{figure*}
	%	\figurenum{1}
	\plotone{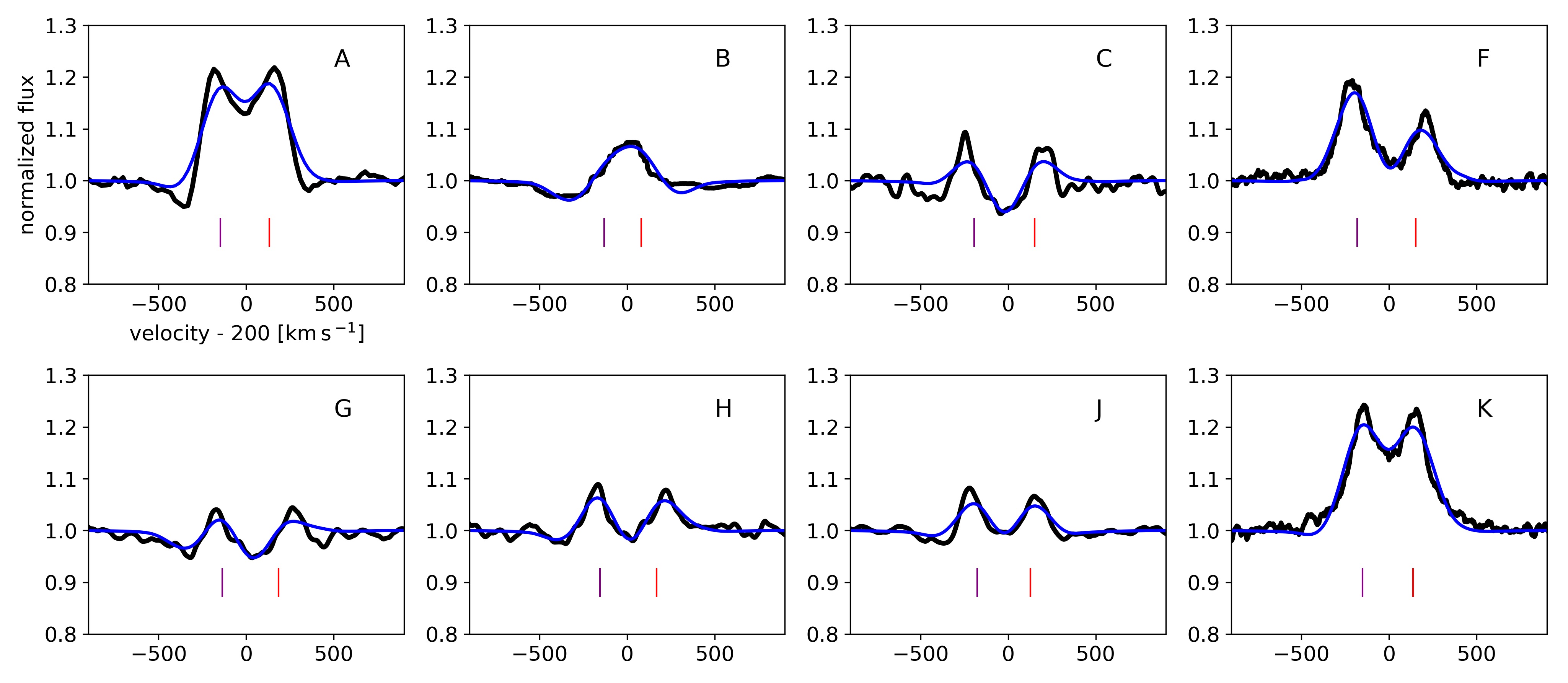}
	\caption{H$\beta$ emission-line profiles from our spectra of AzV~493.  The best-fit photospheric model (Figure~\ref{fig:spec}) is subtracted, after which the violet and red components are fitted by two Gaussian profiles having fixed widths of 2 \AA.  The figure shows the data overplotted by these summed fitted Gaussians.
	The resulting peak values are shown by the vertical lines, and their separations are given in Table~\ref{tab:obs_log}.  Epoch~I has low spectral resolution and is not included in this figure.
	\label{fig:Hbpeak}}
\end{figure*}

Variability in the emission lines is a common characteristic of the Be
phenomenon \citep[e.g.,][]{2013A&ARv..21...69R, richardson:21}.
One effect is the violet-to-red (V/R) variations, which are cycles that can last weeks or decades.  The V/R variations describe changes in the dominant peak strength for double-peaked emission lines observed in some stars.  These cycles are attributed to variation in the morphology and density of the circumstellar disks \citep{1982IAUS...98..453P,1991PASJ...43...75O}.

Figure~\ref{fig:Hbpeak} shows H$\beta$ profiles in the spectroscopic epochs where this line is available, and Gaussian models used to disentangle the V and R components.
The two peaks are clearly resolved in all our observations of H$\beta$, except for Epoch B, which instead shows a P-Cygni profile (Figures~\ref{fig:all}, \ref{fig:Hbpeak}; see Section~\ref{sec:epochBF}).  
Table~\ref{tab:obs_log} gives the peak separations $\Delta\rm H\beta$ fitted in Figure~\ref{fig:Hbpeak}.
The V peak is usually more prominent than R.
{There may be a long-timescale V/R cycle, but further spectroscopic monitoring is needed to determine whether V/R indeed oscillates, or whether there is any trend in $\Delta\rm H\beta$ with phase.
}

\subsection{ Epochs A and K: Evidence of disk evolution}\label{sec:epochA}

Epoch A is observed at a phase of 0.54
{(0.08), soon}
after the apparent eruption event in 2009
(Figure~\ref{fig:ogle_LC}, Table~\ref{tab:obs_log}).
This spectrum shows the strongest helium line emission (Figure~\ref{fig:all}), although we have no other spectroscopic observations within several years of this data point.  Only photospheric \heii\ is seen in absorption in this spectrum; the \hi\ and \hei\ lines are all in emission or filled in.  Moreover, \heii\ $\lambda$4686 is also in emission, which prompted  \citet{2016ApJ...819...55G}, to identify this spectrum as the hottest-known observation of the OBe phenomenon.  
Nebular \heii\ is only generated by the very hottest O stars \citep[e.g.,][]{Martins2021}.

All of the emission lines in Epoch A are double peaked.  \hb\ and
\hg\ show larger peak separations than the \hei\ and \heii\ emission
lines.  For a Keplerian disk, this would imply that the
higher-temperature species is dominated by larger radii than the
\hb\ and \hg\ emission.
Figure~\ref{fig:all} shows that the emission is slightly redshifted relative to the photospheric Balmer absorption.

{
Epoch~K, observed at phase 0.37 (0.74)
(Figure~\ref{fig:all}; Table~\ref{tab:obs_log})
shows the opposite relation between ionization and disk radius.}
Here, the \hei\ lines have larger peak separations than \hb, implying
that the hotter species dominates at smaller radii, unlike Epoch~A.
We also see that the \hb\ and \hg\ line profiles show high-velocity
wings that are not observed at other epochs,
{consistent with high-velocity gas at smaller orbital radii.
Epoch~K is similar in emission-line strength }
to Epoch A and shows \hei\ in emission, but \heii\ $\lambda$4686 is in absorption in this observation, as it is in all the other observations of this line.

\begin{figure*}
	%	\figurenum{1}
	\plottwo{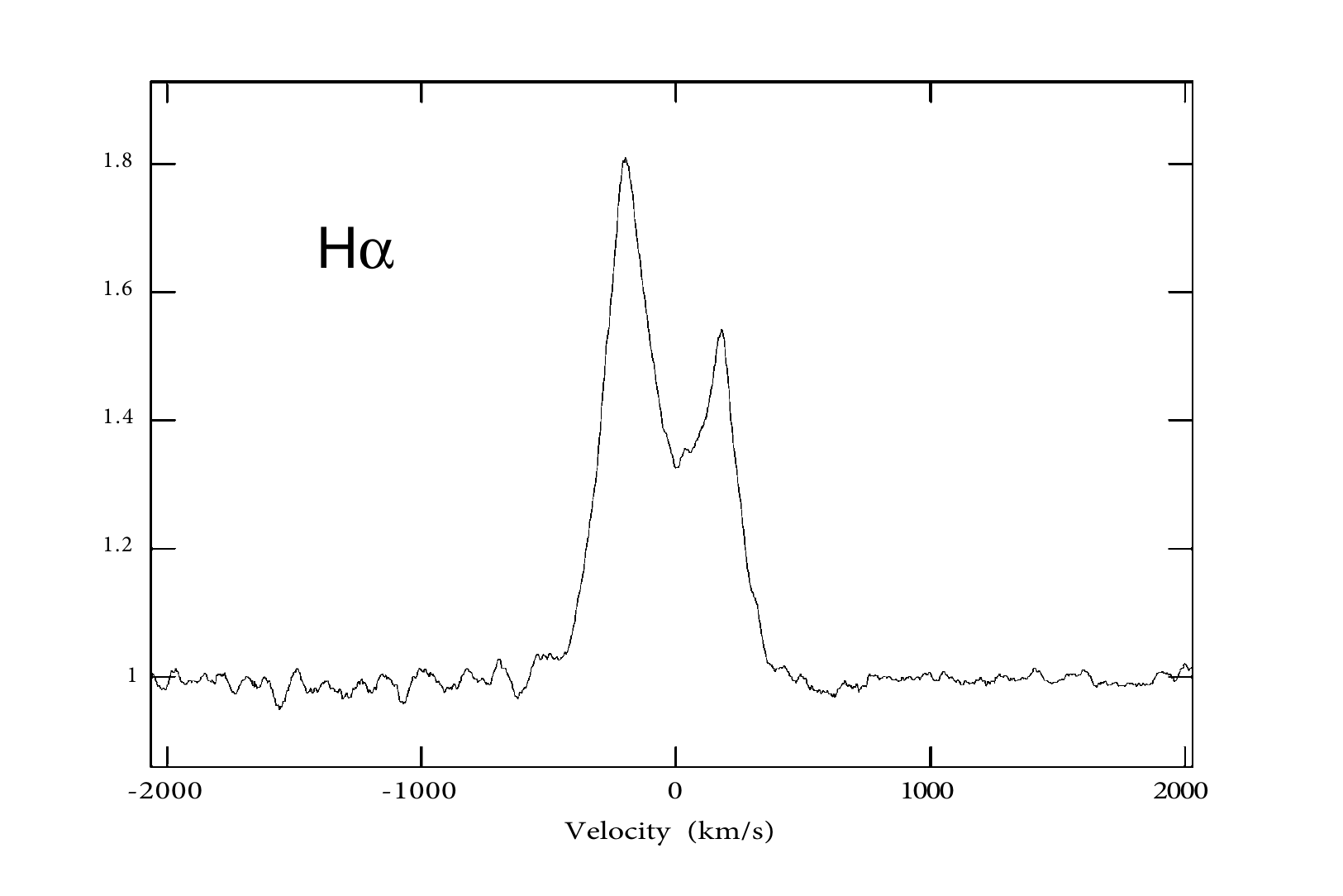}{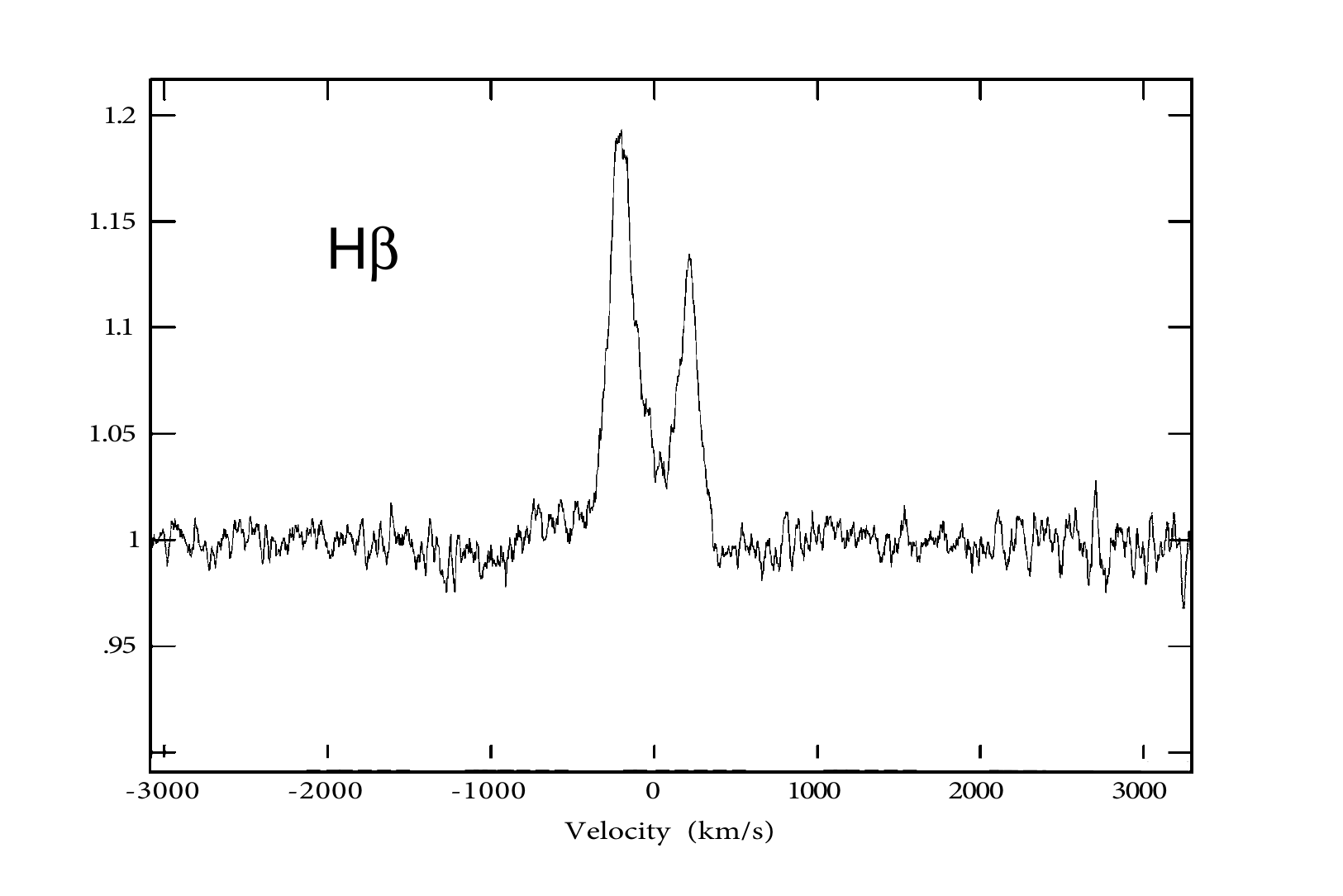}
	\plottwo{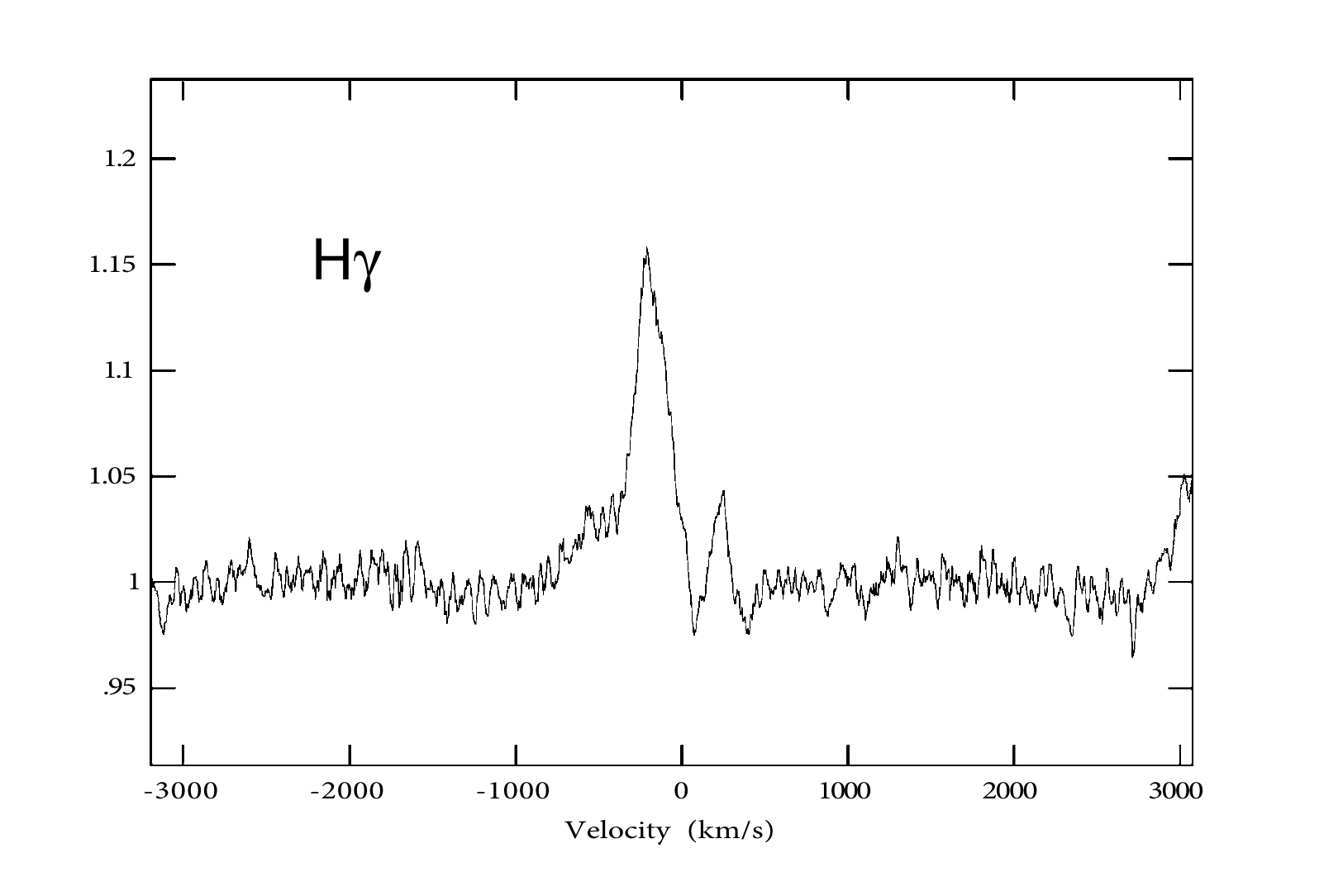}{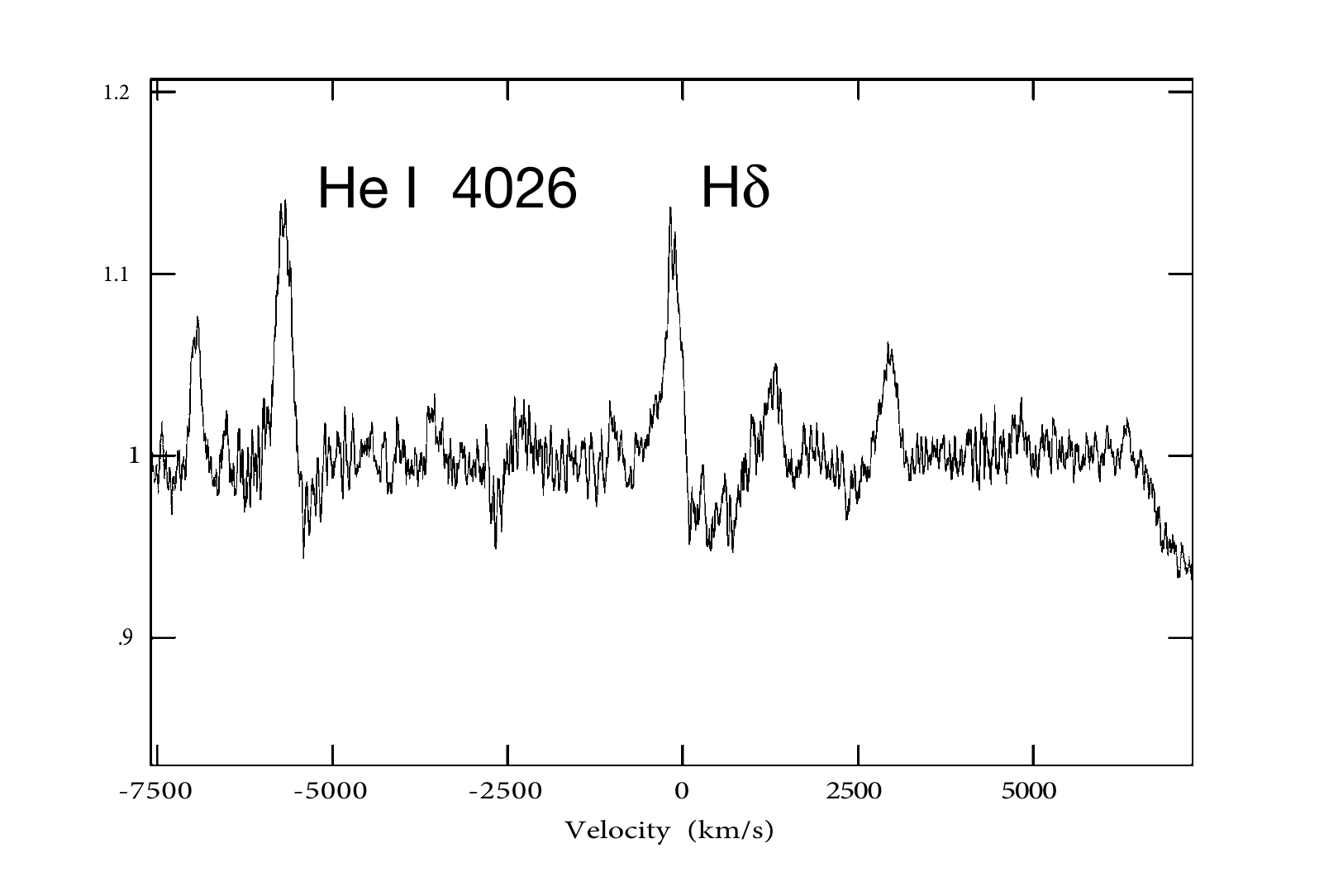}
	\includegraphics[width=3in]{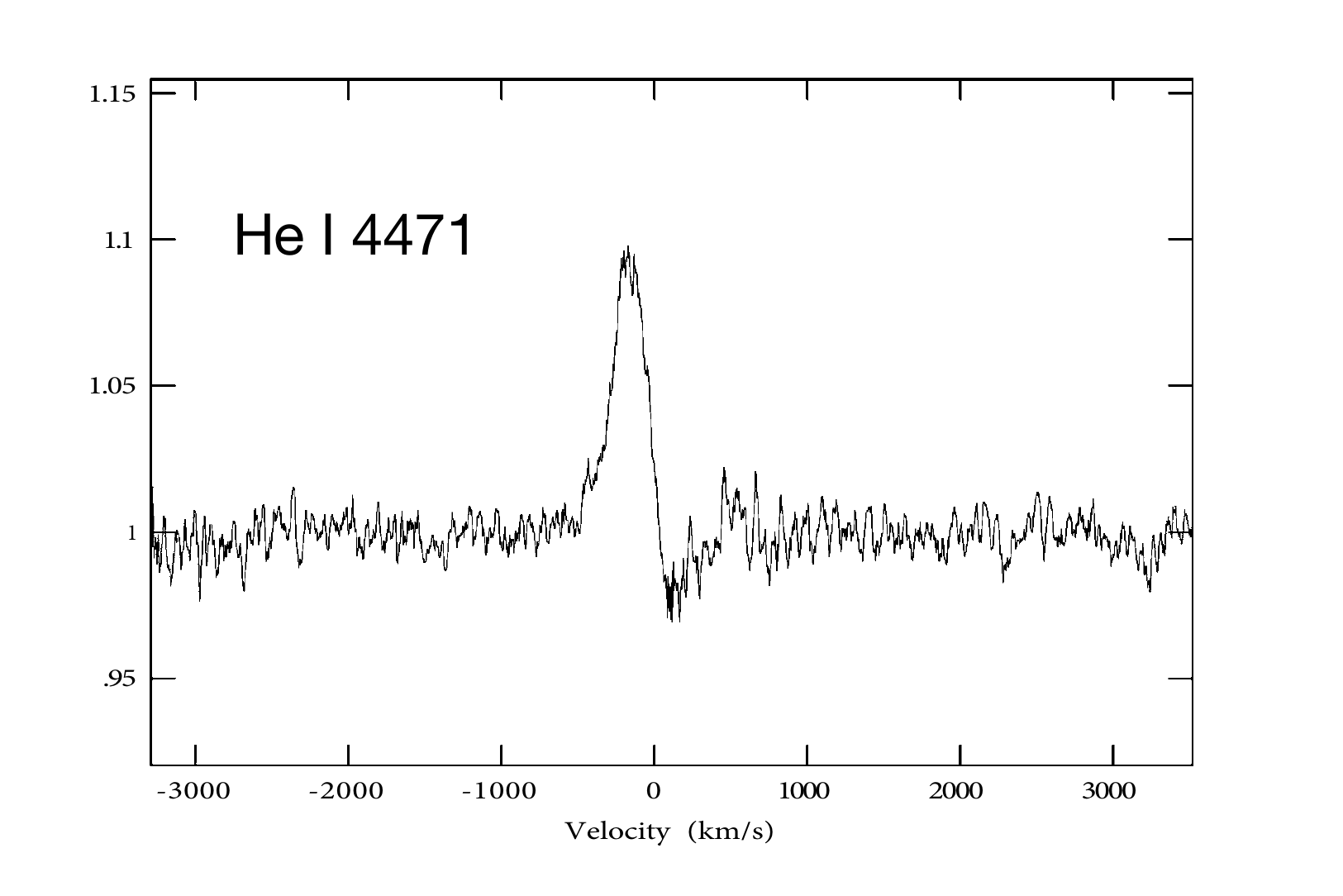}
	\caption{Epoch F line profiles for Balmer and \hei\ emission lines, as shown, centered at the systemic velocity obtained from the \heii\ absorption.  This Magellan/MIKE observation was obtained on 2016 September 22 (Table~\ref{tab:obs_log}).
	\label{fig:epochF}}
\end{figure*}

\subsection{Epochs B and F: Gas Outflow and Infall}\label{sec:epochBF}

Epoch B shows P-Cygni emission-line profiles in H$\beta$ and H$\gamma$
(Figures~\ref{fig:all}, \ref{fig:Hbpeak}),
{suggesting an outflow episode.}
This is also the only spectrum obtained during the period where the strong pulsations
dominate the flux (Figure~\ref{fig:ogle_LC}),
{and it is observed at the latest phase, 0.91 (0.82).}
Figure~\ref{fig:GLSfits} shows that the observation coincides with a local minimum in the light curve.
Thus the P-Cygni features could suggest that the pulsations may be
directly linked to mass ejection,
{since it coincides with the star reaching its smallest radius.}

The spectrum of Epoch~F is dramatically different from most of the other spectra (Figure~\ref{fig:all}).  It shows strong, asymmetric Balmer and \hei\ emission that show remarkable, {\it inverse} P-Cygni line profiles, with red-shifted absorption and blue-shifted emission.  Figure~\ref{fig:epochF} shows the line profiles relative to the systemic velocity of the \heii\ photospheric lines.  
Such observations are usually interpreted as infall of matter \citep[e.g.,][]{Hartmann2016},
which appears to imply a re-absorption of decretion disk material.  The free-fall velocity at the stellar surface for our adopted stellar parameters (Section~\ref{sec:stellparams}) is $\sim800\ \kms$, which is consistent with the red edge of the absorption trough seen in \hd\ and \hei\ $\lambda4471$.
The Balmer emission-line intensities do not follow the Balmer
decrement and are almost uniform (Figures~\ref{fig:spec} and
\ref{fig:epochF}), indicating optically thick emission.  This suggests
that the infalling material is also likely dense,
{and thus has high emissivity}.

Although Epochs D and E are taken only 14 and 11 days before Epoch F,
respectively, Epochs D and E show most lines in absorption with no
sign of these features.  Similarly, Epoch G is obtained only 73 days
after Epoch F, and also shows primarily an absorption spectrum.  Thus,
{this infall episode corresponds}
to a short-lived event, which we fortuitously captured with this MIKE observation.
In the spectra observed before and after Epoch F, the Balmer emission, which presumably originates from the disk, does not seem substantially different in intensity.  This suggests that the reabsorbed material corresponds to a negligible fraction of the disk mass.
The timing of Epoch F is
{at a very early phase, 0.024 (0.05), only}
27 days after the light curve minimum on MJD 57626.
There is no significant feature in the photometry near the time of
Epoch~F, and the light curve is  gradually brightening during this
phase.
{This similarly implies that the continuum luminosity is dominated by
the star and/or disk sources unrelated to the P-Cygni event.}

\section{Disk Ejection Scenario}\label{sec:model1}

The distinctive shape of the light
curve seen in 2002 -- 2004, and again in 2016 -- 2018, showing
a strong drop in brightness followed by gradual increase (Figure~\ref{fig:ogle_LC}),
is seen in some other emission-line stars \citep{2013A&ARv..21...69R}.
We {suggest that this may be due} to the repeated ejection of 
an optically thick circumstellar decretion disk,
{perhaps related to interaction with a binary companion}.
The exact reproduction of this part of the light curve across two cycles, starting with a 1.2-magnitude drop in
brightness, suggests a geometric extinction effect caused by an optically thick disk.
{This event's pattern in photometry and \hb\ line profile is consistent
with a disk ejection outburst, similar to, e.g., HD 38708 \citep{Labadie2017}.}

Assuming that an optically thick disk is indeed expelled to
generate the deep light-curve mimima ($I\sim 14.85$) in 2002 and 2016,
we can estimate the geometric obscuration by considering the maximum
flux following these minima, which peaks around $I\sim 14.0$.  The
difference of 0.85 mag corresponds to reduction in flux by a factor of $\sim 0.46$, or over half, assuming that all of this difference is due to obscuration.  This suggests not only a fairly high angle of inclination, but also a thick, or in particular, a geometrically flared disk, which is consistent with spectroscopic evidence (Section~\ref{sec:epochA}).

{In this model, most of the emission lines originate from an inner
  disk region that experiences variable obscuration to our line of
  sight from a thicker outer disk or torus.
The weaker spectroscopic epochs in Figure~\ref{fig:all} with the 
typical OBe spectrum are the most obscured,
while Epochs~A, B, and K are less obscured and therefore show stronger emission-line spectra.
Epoch C is observed in 2016
at a phase of 0.01 (0.01), and thus near the same phase as the light curve
peak in late 2001 (2009)}
(Figure~\ref{fig:ogle_LC}; Table~\ref{tab:obs_log}).
However, as noted above (Section~\ref{sec:phot}), although the light
curve repeats the disk ejection pattern, there is no evidence of a
corresponding peak preceding this sequence on the same timescale as
that in 2002.  The Epoch~C
{\hb\ profile (Figure~\ref{fig:Hbpeak}) is}
consistent with the optically thick disk already having formed.
Epochs D and E, observed immediately after this minimum, are similarly
unremarkable, although they cover a much shorter spectral range.
Since we see that a putative disk ejection apparently occurred in
2016, it may be that the system has precessed such that an associated
photometric outburst is obscured by the ejection process. 

{The emission lines in Epoch~A are dominated by higher temperature
species at larger radii, whereas Epoch~K shows the opposite effect  (Section~\ref{sec:epochA}).  Epoch~A}
is consistent with very dense, optically thick disks that have
extended vertical flaring, as shown in models by, e.g.,
\citet{Sigut2009}, where the emission, including from harder
radiation, is dominated by this outer region.
{In contrast, the disk geometry at Epoch~K} is dominated by high-density gas near the center and no flaring, thus differing significantly from Epoch~A.  
{Epoch~A is observed at a phase of 0.54 (0.08), and Epoch~K shows the system at a phase of 0.37 (0.74; Table~\ref{tab:obs_log}, Figure~\ref{fig:ogle_LC}).
This suggests that the disk changes between}
having a large, flared outer region at Epoch~A that contributes significantly to the emission, and a configuration 
where flaring is insignificant and emission is dominated by a dense central region
at Epoch~K, perhaps also reaccreting material onto the star.
{The existence of two different components dominated by inner and outer
regions, respectively, could also be due to disk tearing, resulting in
an inner disk and outer, expanding annulus with different inclinations
\citep{Suffak2022, Marr2022}. }

The decreasing \hb\ peak separations seen from Epoch C (346 \kms) to
Epoch J (303 \kms) and to Epoch K (289 \kms; Table~\ref{tab:obs_log})
suggest that the emission is weighted toward increasing radii over
this period, which is consistent with the inner disk dissipating or
{forming an annular disk with an expanding inner radius.}
However, this scenario does not
explain the strong line emission in Epochs A and K
(Figure~\ref{fig:all}), which have the minimum \hb\ peak separations.  If
the inner radius is indeed expanding, then the emitting region either
must become dense, or the disk must precess to lower inclination
angles to reveal stronger line emission.   
The latter could also contribute to a model in which the decreasing
peak separation is due to decreasing obscuration of the disk, allowing
emission at larger radii to dominate.  This
{is consistent with } the system's increasing brightness over this
period (Figure~\ref{fig:ogle_LC}). The extinction may result from the
outer component, or optically thick torus or flare in the disk which either
precesses or dissipates.  However, we caution that such a fast
precession rate may not be feasible.
{Moreover, if the long-term photometric cycle is due to
  precession, the light curve should be symmetric around the minima,
  whereas the observed strong, sudden drops (Figure~\ref{fig:ogle_LC}) are difficult to explain
  with such a model.}

{The outflow and inflow episodes described in
  Section~\ref{sec:epochBF} apparently are not significant in mass relative
  to the entire disk.}
If the minima of the 14-year light curve indeed correspond to the bulk of disk ejection, followed by gradual disk dissipation, then the mass ejection associated with the P-Cygni features in Epoch~B are not likely to be a dominant source of disk material.  However, we note that pulsations have been suggested to be important in replenishing the disk in other OBe systems \citep[e.g.,][]{Baade2016, Baade2018}.

The timing of Epoch F is 27 days after the light curve minimum on MJD 57626.  Although there are 3 other intermediate spectroscopic epochs between the putative disk ejection and Epoch~F, this still takes place during what we assume is the heavily obscured phase in the light curve.  
The lack of any photometric event near the appearance of inverse P-Cygni features in epoch~F suggests that the reabsorbed material is an insignificant portion of the disk material.
The disk is therefore substantial and can plausibly provide material
that may fall back to the star.  {This is consistent with the optically
thick conditions indicated by the Balmer decrement in Epoch~F.}

{Thus, this model is driven by}
repeated ejection of a flared, optically
thick disk whose outer region gradually dissipates, revealing the inner, line-emitting region.
A flared disk is most clearly implied by the ionization and emission-line peak separation in Epoch~A (Section~\ref{sec:epochA}), and
is also consistent with a {\bf maximum} geometric obscuration that may
be $> 50$\% implied by this model.
{The spectroscopic variation could also be caused by disk tearing
  or precession of the system.}
The decreasing trend in \hb\ peak
separations with increasing flux suggests that more light from larger
radii can be seen (Section~\ref{sec:VR}).
Additionally, the high-amplitude, semi-regular pulsations with the $\sim$month-long
period become visible at low extinction (Figure~\ref{fig:ogle_LC}).
Other photometric and spectral variations may be due to contributions
from the inner disk's radial expansion, reabsorption, or
evaporation/ionization, and possible geometric distortion or warping
of the disk system.  

\section{Disk Growth Scenario}\label{sec:diskdiscussion}

{However, some observations seem inconsistent with a disk ejection
model.  For example, the system is bluest when faintest
(Figure~\ref{fig:ogle_LC}), contrary to expectations for reddening.
As noted above, the strong emission-line spectra at Epochs~A and K seem
inconsistent with a dissipating inner disk scenario implied by the
trend in $\Delta\rm H\beta$.  If the long-period cycle is attributed
to disk precession, it would require an additional mechanism to
explain the assymmetric light curve, and also a third, external
massive star that is not seen, to torque the disk.  Thus, alternative
models for the \azv\ system should also be considered.} 

Some other Be stars such as $\delta$~Sco \citep{Suffak2020} and
$\omega$~CMa \citep{Ghoreyshi2018} show long-term photometric
variability in which the increasing flux is due to contributions from
a growing disk, while the minima represent episodes of disk
destruction by the secondary at periastron.  Such a model is therefore
opposite to
{the one presented above.}
In this alternative scenario, the light curve minima of \azv\ in 2002 and 2016 (Figure~\ref{fig:ogle_LC}) correspond to episodes with the lowest disk contribution.  The disk then grows and brightens, recovering its full size around 2005.  In this case, the decreasing trend in \hb\ peak separation with increasing flux is simply due to the disk growth itself.
This scenario is consistent with the blue color at the light curve
minimum in 2016 (Figure~\ref{fig:ogle_LC}), and the weak emission-line
spectra near the 2016 minimum (epochs~C -- J; Figure~\ref{fig:all}).

\begin{figure}
	%	\figurenum{1}
\plotone{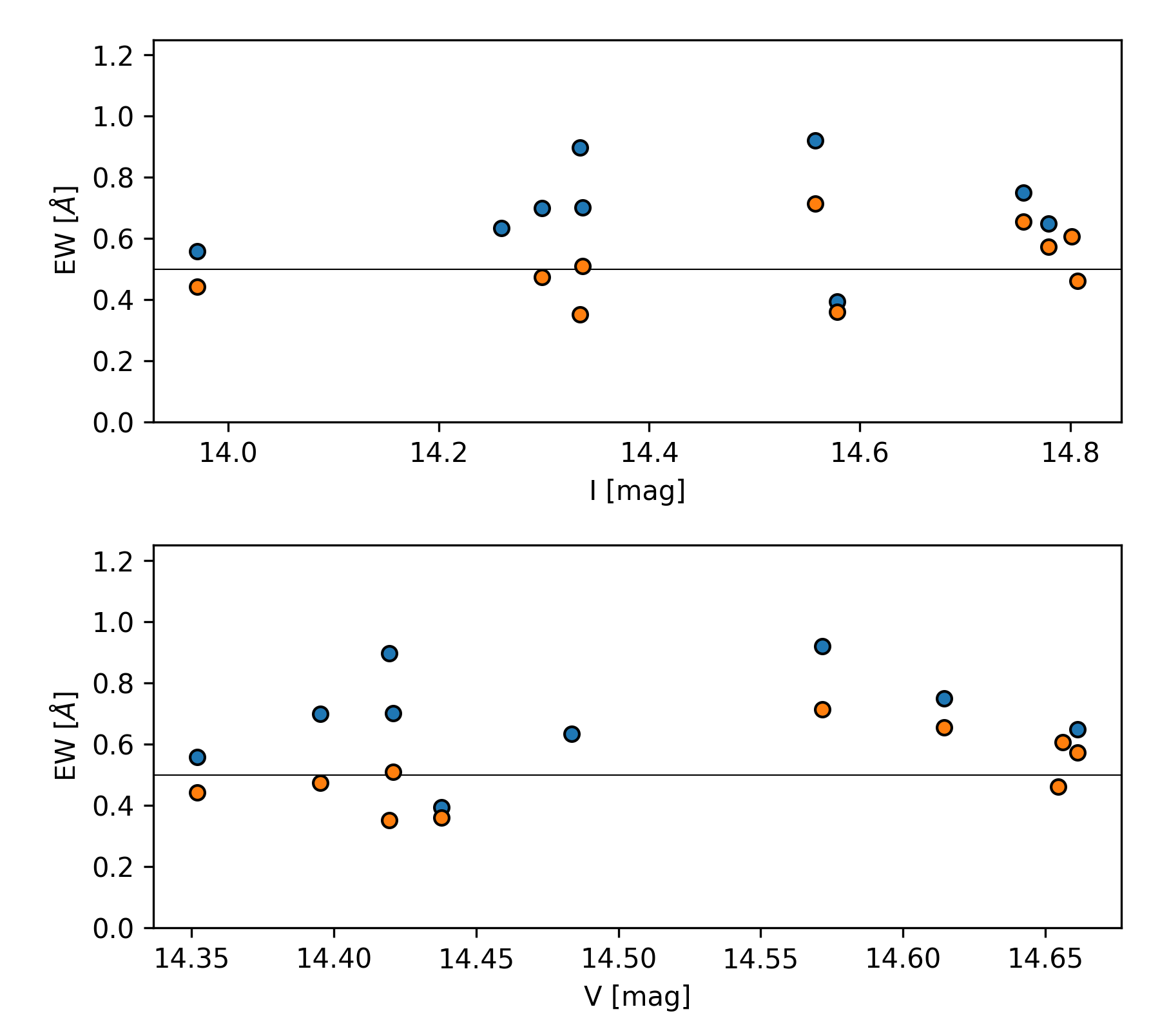}
\caption{{
Equivalent width of \heii\ $\lambda4200$ (red) and $\lambda4540$
(blue) as a function of $V$ (bottom) and $I$ (top) magnitude.  A constant value of
\new{$0.5$ \AA\ }
is shown for reference.}
          \label{fig:EWHeII}}
\end{figure}

If the disk is responsible for the factor of 2.2 increase in flux,
then the equivalent width (EW) of stellar absorption features should
decrease proportionately.  Figure~\ref{fig:EWHeII} shows the EW of
\heii\ $\lambda4200$ and $\lambda4540$ as a function of $V$ and
$I$ magnitude.  A slight trend is indeed apparent, although not as large
as a factor of two in amplitude.
\new{These lines are in the $B$-band, and thus not in the range of our
  photometry.  Figure~\ref{fig:ogle_LC} shows
  that the amplitude of the photometric variations may be smaller at
  bluer wavelengths, although with the given $V$-band sampling it is
  not entirely clear.}
It may be challenging for the
alternative model to produce and maintain the viscous disk necessary
to generate continuum luminosities that compete with those of the
star, given the harsh circumstellar environment of an extreme,
early-type O-star.  

The extinction-dominated model is supported by the lack of correlation
between the strength of the emission-line spectrum and photometric
flux from the system.  There is no significant variation between
spectral Epochs~C -- J (Figure~\ref{fig:all}), which should correspond
to the period of strong disk growth in this model, whereas
the obscuration-dominated model implies dissipation (Section~\ref{sec:epochA}).  The one
exception showing spectral variation, Epoch~F, has P-Cygni emission and stronger emission-line
features, yet it is photometrically unremarkable (Section~\ref{sec:epochBF}).   
Another issue is that the photometric minimum corresponds to the bluest color (Figure~\ref{fig:ogle_LC}), which is more consistent with the alternative model.  However, the star itself may be changing substantially in magnitude and color.
Blueing is also caused by scattering in high-extinction conditions, as seen in the UXOR class of Herbig Ae stars \citep{Natta2000}. 

The overall shape of the light curve for \azv\ is rather different
from those of $\delta$~Sco and $\omega$~CMa, which show extended
minima with more top-hat-like light curves \citep{Ghoreyshi2018,
  Suffak2020}.  In contrast, \azv\ shows sharp minima
(Figure~\ref{fig:ogle_LC}), implying very rapid disk destruction and
immediate, regular regrowth in the alternative model.  It seems hard
to explain such sudden dissipation of a several-AU dense, viscous disk
by a  neutron star or black hole (Section~\ref{sec:binary}) during the
brief periastron passage.
{Moreover, the exact reproduction of the photometric cycle's
initial segment (Section~\ref{sec:phot}) is unusual and may be harder
to explain with a disk-growth model.}

Overall, the fundamental nature of the light curve and disk evolution remain unclear.
Tailored modeling of this system and further multimode observational monitoring is needed to clarify the relationship between the decretion disk and interaction with a secondary star.

\section{An Extreme Interacting Binary}\label{sec:binary}

The fast surface rotation for
  this evolved O star is a natural signature of accretion during a mass
  transfer event \citep[e.g.,][]{packet:81, cantiello:07,
    renzo:21zeta},
consistent with an interacting binary scenario.  {If}
the disk is induced by a periastron passage of an undetected
companion,
{then this may imply a long, 14.6 (7.3)-year period, and hence
a large and highly eccentric orbit.
For the \azv\ stellar parameters obtained in Section~\ref{sec:stellparams}, 
a neutron star companion of mass $1.4\,M_\odot$ would require
$e\sim 0.95$ (0.93) and apastron of
$\sim 43$ (27) AU} for a typical OBe star periastron distance of
40$R_\star$.
{These orbital parameters are similar to those of the Be star $\delta$ Sco}
\citep[e.g.,][]{Che2012}.
The unseen companion could also be a somewhat more massive main-sequence or stripped star, or a black hole.
{The eccentricity may be lower, but if a binary companion is
  responsible for disk ejection, then periastron must be small and the
  eccentricity high.  The nominal periastron value used here would likely be
 an upper limit, since $\delta$ Sco showed no disk ejection at
 periastron \citep{Che2012}.}

\subsection{Neutron star or black hole?}

{Thus, if a binary companion excites disk ejection or is
  otherwise responsible for the observed properties of \azv, then it
 is probably an eccentric system, and}
the most likely explanation for such an orbit is that the
companion has already experienced core collapse, receiving a strong kick.
Large natal kicks are routinely invoked in core-collapse events that form neutron stars \citep[e.g.,][]{arzoumanian:02, podsiadlowski:04, verbunt:17b, Janka2017}. Natal kicks during black
hole formation are still highly debated \citep[e.g.,][]{dray:05,janka:13, mandel:16, repetto:17, atri:19, renzo:19walk, callister:20}, but not excluded.
Assuming a large $450\,\mathrm{km\ s^{-1}}$ kick, \cite{brandt:95} found a broad
correlation between eccentricity and orbital period of binaries
surviving the first core-collapse. This is in agreement with the
high $e$ and long period we find for \azv. 

The present-day mass of \azv\ can be used to constrain
the nature of {a putative} compact object. Assuming a flat distribution in
  initial mass ratio, the average initial binary mass ratio
  $q=M_2/M_1\simeq0.5$ \citep[e.g.,][]{Moe2017}. Without any accretion during mass
  transfer, the present-day mass of \azv, $M_2\simeq50\,M_\odot$, would
  suggest $M_1\simeq100\,M_\odot$, which at SMC metallicity
implies that the compact object should be a black hole
\citep[e.g.,][]{sukhbold:16, couch:20, zapartas:21}. In this case, however, the rapid rotation
of \azv\ would need to be primordial.
  
Instead, it is more likely that
mass transfer has occurred, in which case
$M_1$ is likely to be quite different, depending on the mass transfer efficiency. A mass transfer
phase during the donor's main sequence (Case A)
  is expected to be slower and more
  conservative, possibly causing significant mass growth of the accretor without
  extreme chemical pollution.  This scenario has been
  invoked to explain the formation of low-mass compact objects in very
  young regions \citep{belczynski:08}, and in particular, the origin of very massive companions
  \citep{vandermeij:21}, such as we have for \azv.
In this case, the zero-age-main-sequence (ZAMS) mass of $M_1 \sim 30 - 40$ \msun\ for the adopted $q$, also accounting for the final donor core mass. However, mass transfer is far more likely to occur after the donor main sequence (Case B), due to the star's expansion \citep[e.g.,][]{vandenHeuvel1969}.  It then takes place rapidly, on the thermal or He core-burning nuclear timescale \citep{Klencki2022}, and system mass loss is far more likely, implying a higher ZAMS mass for $M_1$.

Although post-SN outcomes are stochastic, black hole production is
expected to dominate for $Z_\odot$ progenitors with initial masses
$\gtrsim 20$ \msun. This nominal threshold ZAMS mass is expected to
decrease for lower metallicity
\citep[e.g.,][]{zhang2008,oconnor2011,sukhbold:16}, which in principle
enhances the likelihood that the compact object should be a black
hole. The high eccentricity in AzV 493 strongly suggests that a SN
occurred. While this implies that the companion is more likely to be a
neutron star, black holes can form from fall-back if the SN is
insufficient to unbind the ejecta, which is more likely to happen at
low metallicity \citep[e.g.,][]{zhang2008}. There are multiple
mechanisms to produce core-collapse black holes, and if mass-loss
occurs, a SN and/or kick to the system may result
\citep[e.g.,][]{janka:13}. We note that $M_1 \sim 20 - 40$ \msun\ is a
range that has been extensively simulated and where explodability and
fallback are uncertain
\citep[e.g.,][]{oconnor2011,sukhbold:16,janka:13,zhang2008}. Establishing
that a neutron star or black hole resulted from this ZAMS range, with
some kind of kick, would provide an important empirical reference for
theoretical models of the explosion and the binary interactions
preceding it.  

Follow-up observations at subsequent periastra could more firmly
establish whether \azv\
{has a companion, and whether it}
is a black hole vs a neutron star.
A 74.33 ksec Chandra/HRC observation on 2012 February 12 (MJD 55969) of a field including \azv\  (ObsID 14054) shows no detection.
Given the tiny orbital interval during which the two stars interact, no significant accretion onto the compact object is expected, explaining why the system is not a known high-mass X-ray binary.  However, well-timed X-ray observations near periastron may be able to catch a brief flare event.

\subsection{Radial Velocities}\label{sec:RV}

\begin{figure*}
%	\figurenum{1}
\plottwo{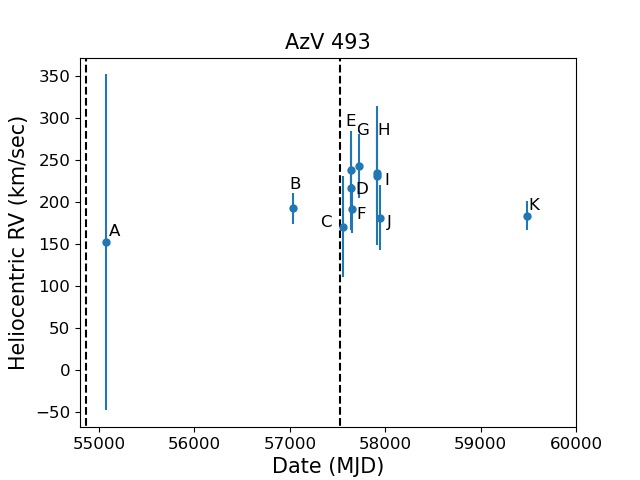}{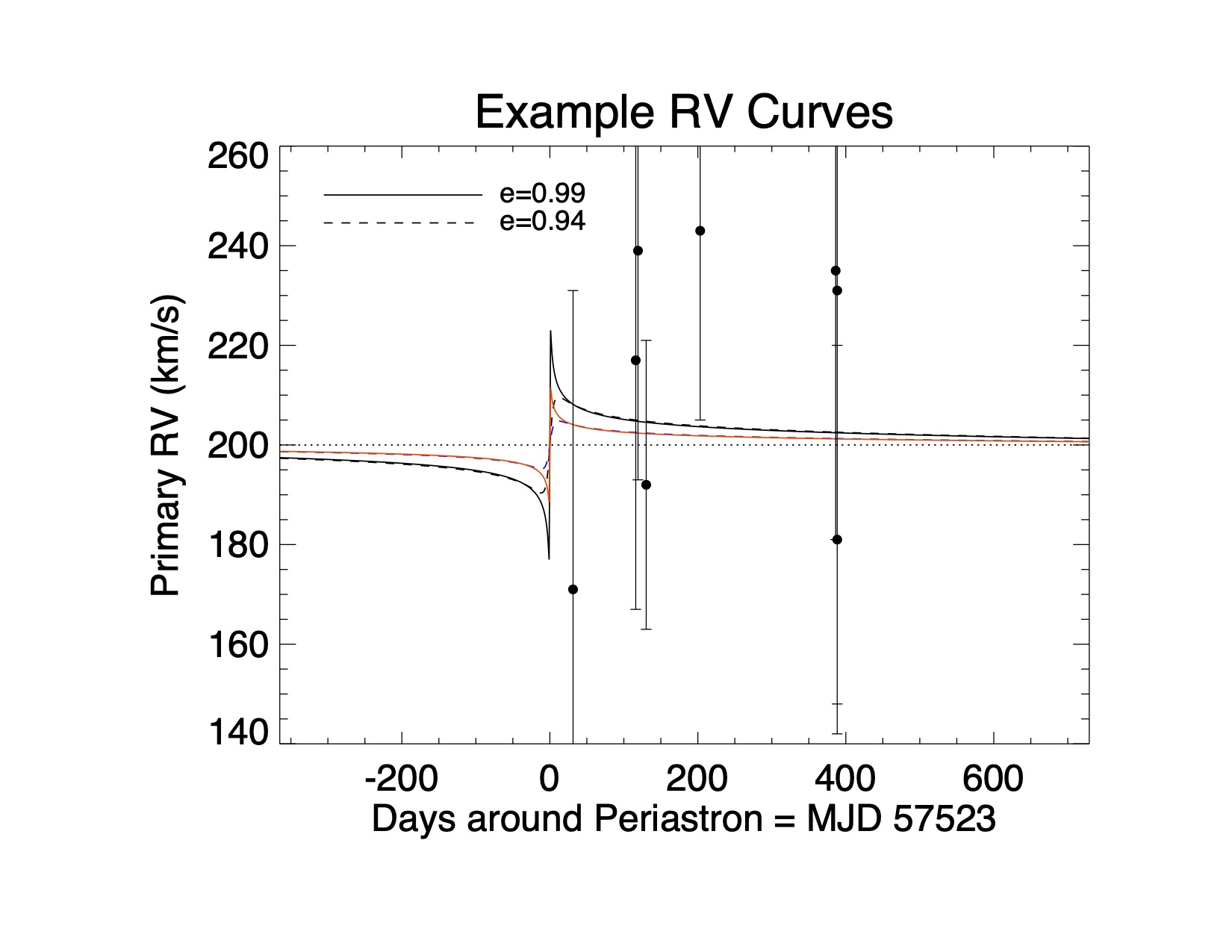}
\caption{Left: Heliocentric radial velocities measured from
 \heii\ photospheric absorption vs MJD for all epochs.
 Epoch~A has only one available line of low quality, and
 hence has a very large uncertainty.  The vertical dashed lines show the
{possible} periastra at MJD~54686 and 57523.
Right: Zoom for the same data showing RV models for
eccentricities of 0.93 (dashed lines) and 0.99 (solid lines),
{assuming a periastron occurs at MJD~57523;}
and for inclination angles of $90^\circ$ (black)
and $45^\circ$ (blue), for the 50 \msun\ primary and
assuming a 3 \msun\ secondary.  
{If a periastron is closer to the light curve minimum at MJD 57626,
the models would shift to 103 days later.}
          \label{fig:RV}}
\end{figure*}

We also measure the radial velocity (RV) for the obtained spectra
{to search for evidence of a companion}.
This is challenging, since \azv\ is a luminous, fast-rotating,
early-type O-star, with few photospheric features, several of which
are often in emission.  We carried out cross-correlations against the
{\sc FASTWIND} model spectra for the entire observed spectral range
using the iSpec code \citep{Blanco-Cuaresma2014}, as well as
determinations based on cross-correlations against PoWR model spectra
\citep{Hainich2019} for only the \heii\ lines ($\lambda4200$, $\lambda
4540$ lines, and $\lambda4686$), which are the only clean features
appearing in all epochs.  The latter are carried out with the Markov
Chain Monte Carlo code of \citet{Becker2015}, and since they yield
better results, we adopt these RV measurements (Table~\ref{tab:obs_log}).

We find that the mean systemic radial velocity is $202\pm 9$ \kms, weighted inversely by the errors.  We caution that the quoted standard error on this value underestimates the uncertainty if there is true variation.
Given the difficulty of these measurements, with median error on
individual epochs of 46\ \kms, it is difficult to evaluate any
variability (Figure~\ref{fig:RV}).  There is possible evidence for
very short-term RV variations; however, the data are ambiguous.

{
We compute RV models for a possible periastron 
suggested in Section~\ref{sec:period} at MJD 57523, which is near the second
minimum in the light curve (Figure~\ref{fig:ogle_LC}).  For this 7.3-year period, and
the above, nominal periastron distance of $40R_\star$, the
eccentricity $e\sim 0.93$ and apastron $\sim28$ AU.  For this scenario,}
Figure~\ref{fig:RV} demonstrates that the RV signature of
a neutron-star companion at periastron is
very brief, on the order of 0.01 in orbital phase, and moreover, the
observational uncertainties are larger than the expected amplitude.
{This is the case even for $e = 0.99$.}
Thus, our existing RV measurements do not strongly constrain
{whether MJD 57523 corresponds to a periastron, nor the}
existence and properties of a companion,

\subsection{Proper Motion}

\begin{figure*}
	%	\figurenum{1}
	\plotone{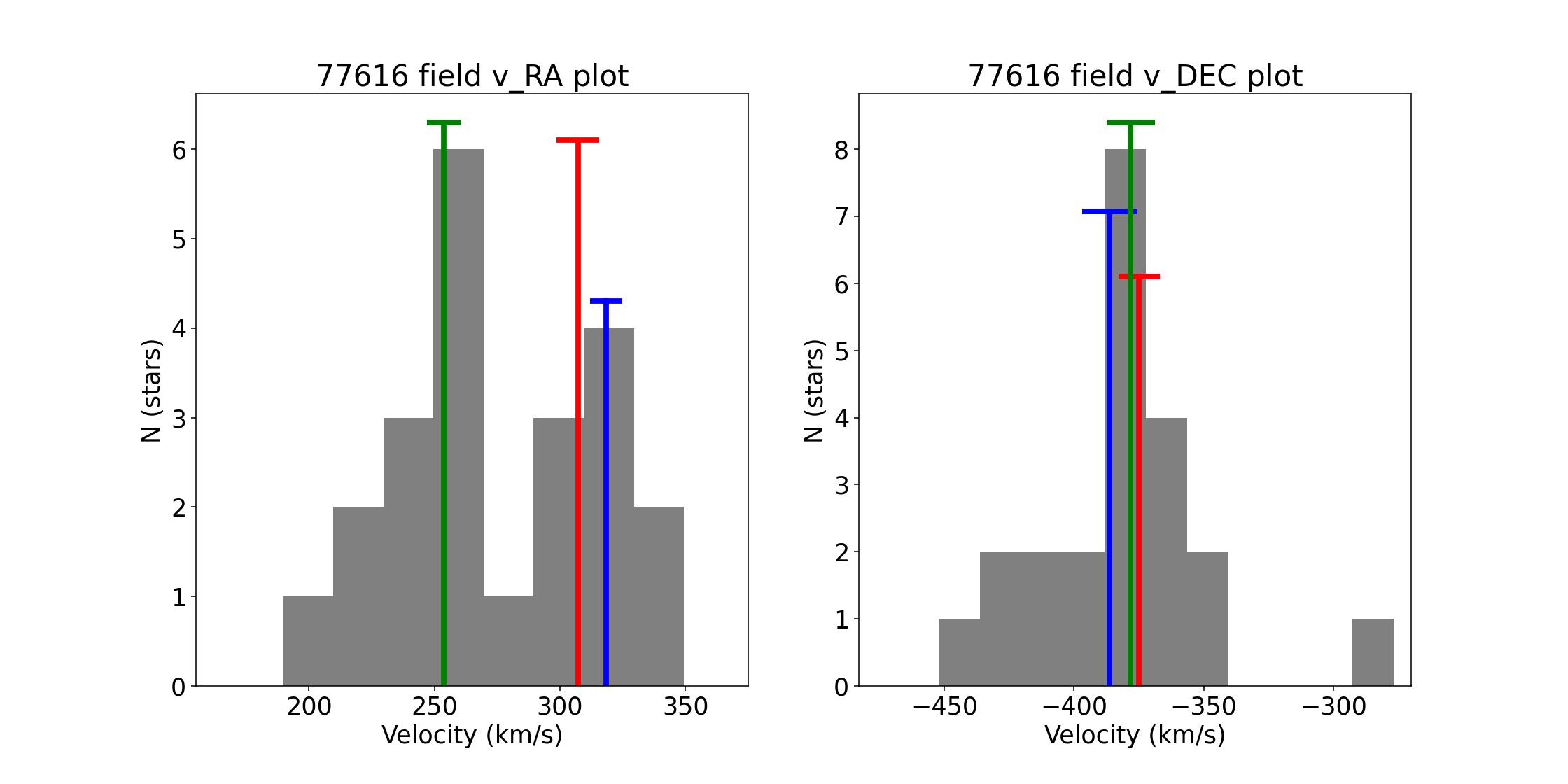}
	\caption{Distribution of Gaia proper motion velocities in R.A. (left) and Dec (right) for stars from \citet{2002ApJS..141...81M} within $5\arcmin$ of \azv.  The bimodal R.A. distribution defines two kinematic groups.  The first group has 13 stars with median velocity $(v_\alpha, v_\delta) = (254\pm7,\ -378\pm9)\ \kms$ and the second has 10 stars with $(v_\alpha, v_\delta) = (318\pm6, -386\pm11)\ \kms$.  The one star between the two groups in $v_\alpha$ is included in both.  The median velocities for these groups are shown with the vertical green and blue lines, together with the velocity of \azv\ (red).
          \label{fig:PMhist}}
\end{figure*}
\begin{figure*}
%	\plotone{FOV30arcmin_radius.jpg}
 	\plotone{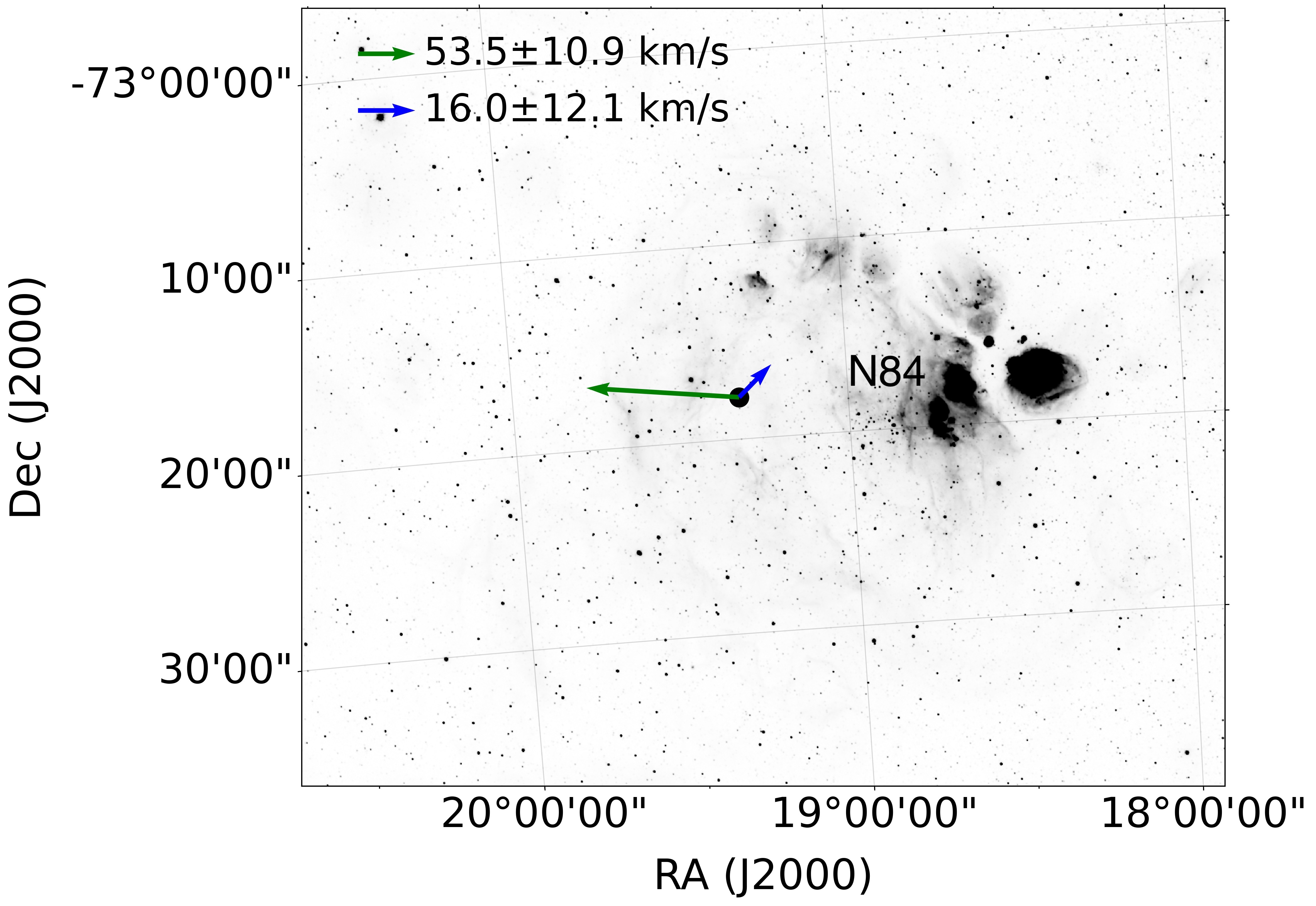}
	\caption{Location of \azv\ in the SMC field, with the green and blue proper motion vectors corresponding to the two field velocities indicated with the same color coding in Figure~\ref{fig:PMhist}, superposed on an H$\alpha$ images from \citet{Smith2005}.  
    The nearest massive star-forming region is the N84 complex \citep{Henize1956}, indicated.  For the adopted SMC distance, $10\arcmin = 181$ pc.
          \label{fig:FC}}
\end{figure*}

A post-SN bound system can be expected to have been accelerated from its original rest frame.
Relative to the blue stars from \citet{2002ApJS..141...81M} within a 5$\arcmin$ radius, the Gaia EDR3 \citep{Gaia2021} residual proper motions of \azv\ show two potential velocity vectors.  Figure~\ref{fig:PMhist} provides the velocity histograms of these local field stars, showing strong bimodality in the R.A. components.  These define two possible local velocity fields implying R.A. and Dec residual velocity for \azv\ of either $(v_\alpha, v_\delta) = (53\pm11, 3\pm12)\ \kms$; or $(v_\alpha, v_\delta)= (-11\pm11, 12\pm13)\ \kms$.  These yield total projected transverse velocities of $54\pm 11\ \kms$ or $16\pm12\ \kms$.

Figure~\ref{fig:FC} shows a wide-field view of the surrounding environment,
with the two possible proper motion vectors superposed.
We see that the nearest massive star-forming region is the N84 complex
\citep{Henize1956} about $15\arcmin - 20\arcmin$ or $\sim 300$ pc to
the west.  If the velocity measurements are correct, the faster,
east-bound velocity is consistent with \azv\ originating in N84 and
traveling for $\gtrsim 5$ Myr.  The lifetime itself of a 50
\msun\ star with \vsini\ $\sim 500\ \kms$ is about 5 Myr
\citep{2011A&A...530A.115B}, and for a SN ejection, its travel time would only be the post-SN
lifetime.  However, since the star presumably acquired its total mass and spin
later in life, the system may have been ejected earlier by
dynamical processes as a tight, non-compact binary.  If so, it would
have been reaccelerated by the SN explosion, therefore implying that
it may be a two-step ejection \citep{Pflamm2010}.  Supernova
accelerations are typically weaker than dynamical ejections
\citep[e.g.,][]{renzo:19walk}, and so the dominant velocity component could
still be due to a dynamical ejection from N84.
{A dynamically active past in a dense stellar
environment of N84 may also help to explain the eccentricity
\citep[e.g.,][]{SimonDiaz2015}, although}
it would seem unlikely that the system could maintain
its highly eccentric configuration for 5 Myr. 
On the other hand, we
note that the inferred runaway velocity, orbital eccentricity, and
period are still consistent with being due solely to SN acceleration
\citep[e.g.,][]{brandt:95}.   
{Thus, in order to explain both the long travel time and high
  eccentricity, the most plausible scenario may be the two-step ejection.}

There is also a small possibility that the
{slow,}
alternative proper motion vector (Figure~\ref{fig:FC}) is correct.
{However, this would mean that
the \azv\ system formed in isolation
since there is no corresponding young cluster whence it could have
originated (Figure~\ref{fig:PMhist}).
\citet{Vargas2020} find that $< 5$\% of OB stars, if any, formed in
the field, and this is especially unlikely
for \azv, given its high mass.  }
We caution that the velocity errors do not include unknown systematic errors, and so these measurements need to be confirmed.  Thus, although \azv\ indeed appears to be a runaway star, this does not provide especially useful information to constrain its binary interaction history.

\subsection{Similar systems}

{A comprehensive study by \citet{Marr2022} shows that the B8 Vpe
star Pleione (HD 23862) has a light curve with a similar long-term
pattern of slow growth with sudden drops, and similar variations in
the Balmer emission-line profiles.  It is a triple system with a close
companion on a 218-day orbit \citep{Katahira1996,
  Nemravova2010}.  \citet{Marr2022} suggest that
the photometric drops correspond to the decretion disk tearing into two
components, where one remains aligned with the star's equatorial plane
and the other is misaligned due to tidal torque from the close companion.
Pleione's long-term photometric cycle is 34 years, similar in
magnitude to that of \azv.  \citet{Nemravova2010} find that the close
companion is on an eccentric orbit with $e > 0.7$.}

\azv's \new{initial} peak brightness and subsequent drop
\new{in 2001 (Figure~\ref{fig:ogle_LC})}
qualitatively
resemble the photometric pattern characteristic of heartbeat stars.  These
are a rare class of interacting binary systems with
high eccentricities such that the periastron passage tidally induces
regular photometric outbursts.
However, the observed pattern in \azv\ cannot be induced by this type of tidal interaction;
preliminary simulations using new capabilities in the GYRE stellar
oscillation code
\new{\citep{Sun2023}}
suggest that the combined amplitude and width of the periastron pulse cannot be reproduced by eccentric tidal models.
\new{Nevertheless, given that \azv\ seems likely to be a massive
  eccentric binary system, massive heartbeat stars thus share some
  similarities with this object if a companion indeed interacts with
  the primary and/or its disk.  Examples include}
the non-Be binary system $\iota$ Ori (O9~III $+$ B1~III/IV), which
has orbital period 29 d and eccentricity $e=0.764$, as determined by
\citet{Pablo2017}.  They find that the two components have masses of
23.2 and 13.4~\msun, respectively, generating tidally excited
oscillations with periods on the order of $\sim 1$ day.  MACHO 
80.7443.1718 is another heartbeat system with two stars of type B0~Iae
and O9.5~V and masses of 35 and 16~\msun, respectively
\citep{Jayasinghe2021}.

{The B0.5 Ve star $\delta$ Sco is has a
B2 V star companion in an eccentric ($e=0.94$) orbit with
period 10.7 years \citep[e.g.,][]{Tango2009, Tycner2011}.  The two
components have masses of 13.9 \msun\ and 6 \msun\ \citep{Che2012}.
This system shows a long-term photometric cycle  somewhat similar to
that of \azv, although much more 
poorly defined.  There is no obvious link between the disk
properties and binary interaction \citep{Suffak2020, Che2012}, but the
long-term photometry has a timescale similar to that of the orbital period.}

H~1145--619 is a Be X-ray binary whose primary is a B0.2e~III star estimated to be 18.5 \msun\ \citep{2017A&A...607A..52A}, and the secondary is an X-ray pulsar.
As shown by \cite{2017A&A...607A..52A}, H~1145--619 has a light curve
with a $\sim10$-year cycle together with unexplained multiple modes of
much shorter periods ($\sim 1$ year), qualitatively similar to
what we see for \azv, which has 
{a long cycle of 14.6 (7.3)} years
and short oscillations of $\sim 40$ days.
{While it remains unclear whether the light curves of H~1145--619 
and \azv\ have fundamental similarities, both stars are massive OBe
stars.  If they are related, the fact that H~1145--619 has a confirmed
compact binary companion may suggest that
the unusual variability of \azv\ may have a similar origin.}

These objects provide a context for \azv\ that supports this object
being a member of this 
{broad class of binary, massive OBe systems with high eccentricities}.
At 50 \msun, \azv\ is more massive than any of these similar objects.
It is also one of the earliest O stars in the SMC, since there is no
photospheric \hei.  Thus, \azv\ may be the most extreme such object known,
in terms of its mass and effective temperature.  Its 
variability amplitudes are also among the largest known. 

We note that, based on only the Epoch A spectrum
(Figure~\ref{fig:all}), \citet{2016ApJ...819...55G} suggested that
\azv\ is a normal, but extremely early, classical Oe star.  Given the
strong spectroscopic and photometric variability, the nature of this
spectrum may be somewhat different than inferred in that work, and the origin
of the strong line emission seen in this particular spectrum is
unclear (Section~\ref{sec:epochA}).  Still, its status as a post-SN
binary where the observed star was likely spun up by mass transfer
from the compact object progenitor, is consistent with the origin of
classical OBe stars.  Indeed, given that most of the massive OBe stars
are post-SN systems \citep[e.g.,][]{Dallas2022,DorigoJones2020}, we
can expect that more of them are likely to be high-eccentricity,
compact-object binaries.

\subsection{Alternative Companion Scenarios}\label{sec:other}

We now consider alternative scenarios for a putative binary component.
First,
{such a}
companion might be an unexploded former donor in an interacting binary.  In this case,
it could be a stripped star \citep[e.g.,][]{schootemeijer:2018, gotberg:17},
which can be elusive to detect. 
\citet{Wang2021} identified hot, stripped star companions to Be stars based on FUV spectral cross-correlations; however, the extremely hot temperature of \azv, which is commensurate with the hottest O stars, poses a serious challenge for this method.
If the observed star has
previously experienced accretion from binary mass transfer, then its surface might be He-
and N-enriched \citep[e.g.,][]{blaauw:93, renzo:21zeta}, although
whether this occurs depends on the accretion efficiency and mixing processes in the accretor's envelope. 
Since early O stars have few metal lines, it is again difficult to evaluate any enrichment, especially in a fast rotator like \azv. There is no immediate evidence for any unusual abundances in this star.
Moreover, a non-degenerate companion does not
naturally explain the high observed eccentricity,
{which would then have to be primordial, avoiding tidal
  dissipation, or of dynamical origin.} 

Alternatively, the high rotation rate and variability of \azv\ might be caused by a
non-standard internal structure of the star because of a
merger. These are common among massive stars, occurring in
$22^{+26}_{-9}\%$ of isolated massive binaries \citep{renzo:19walk},
with an even higher rate if accounting for the presence of further
companions \citep[e.g.][]{toonen:20}.
{For example, $\eta$ Car has been suggested to originate from a
  merger in a hierarchical triple system, resulting in a present-day
  eccentric binary \citep[e.g.,][]{Hirai2021}.  However, $\eta$~Car is
  a luminous blue variable star and has other substantial differences
  from \azv.}

Yet another possibility is that \azv\ might be a triple system with a third, also invisible, star on a shorter-period orbit.  This speculative scenario might help to explain how the strong, 40-day pulsations are maintained (Section~\ref{Sect:Pulsation}).  It also might help explain the apparently sporadic ejection and accretion events seen in Epochs B and F (Section~\ref{sec:epochBF}).  Such a system would be unstable, but the brief interaction phase with the secondary may enhance its longevity.
We note that the system is unlikely to be a triple in which the third star has an even larger orbit than the secondary.  Although high orbital eccentricities can be produced by Kozai-Lidov cycles in such a system,
this high-eccentricity phase of the cycle
is short in duration.
Thus, such extreme eccentricity may require a triple or higher-order
multiple-star interaction in the system's birth cluster, and may be
linked to a dynamical ejection of \azv\ into the field.  Overall,
however, it is challenging to explain \azv\ in terms of a triple-star
scenario.  Unfortunately, RV monitoring is complicated due to the
technical difficulty and
{possible}
presence of varying stellar pulsations, so it will be hard to evaluate whether the system consists of more than two stars.

\section{Summary} \label{sec:sum}

We present 18 years of OGLE Project photometric data and spectroscopic data over 12 years, revealing the remarkable variability of \azv.  This is perhaps the earliest known classical Oe star, with $T_{\rm eff} = 42000$ K, $\log L/L_\odot=5.83 \pm 0.15$, and $R_\star/R_\odot=15\pm3$. These parameters imply a mass of $50\pm9$ \msun.  
The dominant photometric pattern is reproduced after 14.6 years.
There are also large, semi-regular $\sim40$-day pulsations of unknown origin, as well as other structure in the light curve.  It is not a known HMXB.
The observed \vsini\ $=370\pm40$ \kms, with a high inferred $\sin i$,
suggesting a rotational velocity of $400 - 450$ \kms.  The system is
$\sim 300$ pc from the nearest massive star-forming complex and its
proper motion shows that it is likely a runaway star
{from that region, with a transverse velocity of $54\pm11$ \kms,}
possibly having experienced two-step acceleration.   

Altogether, the data suggest that this object is
{likely an eccentric, interacting binary system}
with an undetected compact companion.  If so,
the orbital period could correspond to the 14.6
{(7.3)-year}
period, implying a high eccentricity of at least
{$e\sim 0.95$ (0.93)} and apastron $\sim43$ (28) AU.  
If this binary scenario is correct,
\azv\ would be
{among the most extreme systems known, in terms of its early
spectral type, high mass, and extreme eccentricity.}

In our favored model, an optically thick decretion disk
{is regularly ejected, likely by a periastron encounter.
A two-component disk system forms, with the}
outer region responsible for the 0.85-magnitude drop in $I$-band
flux, while the inner
disk is the origin of most of the observed emission-line spectrum.   
The spectra appear to show varying relative contributions from the
inner and outer regions, consistent with the
{optically thick outer region}
dissipating over the cycle. 
{The outer region may correspond to a flared disk, torus, or
possibly, a separate inclined annulus formed by tearing from the inner disk.}
We see direct spectroscopic evidence for episodes of both matter ejection and infalling
reabsorption of dense disk material onto the star. 
The lack of exact regularity of photometric and spectroscopic variations in the cycle implies that the geometry and/or mechanics of the disk ejection may vary. 
An alternative, opposite model seen in some Be stars, in which the
brightness increases due to contribution from growing disk emission
\citep[e.g.,][]{Suffak2020,Ghoreyshi2018}, should also be considered.   

{If \azv\ indeed has a
highly eccentric orbit, it would suggest} that the system experienced a strong SN kick, implying that the unseen companion is a neutron star or black hole.  The high \vsini\ also suggests that mass transfer occurred before this event.  For conservative, Case~A mass transfer, the progenitor donor's ZAMS mass would be $30-40$ \msun\ for a typical $q\sim 0.5$, and larger for non-conservative Case~B mass transfer.
This mass range is well within that suggested by models to produce
black holes, although the occurrence of strong natal kicks in cases of
black hole formation is less clear.  Alternatively, the donor could be
a stripped star; however, this scenario cannot explain the extreme
eccentricity,
{which would have to be dynamical or primordial.}
The system could also be a merger, but the eruptions
and long-term pulsations seem less consistent with this
scenario. \azv\ could possibly be a triple system, which might explain
how the strong photometric oscillations are maintained (Section~\ref{sec:other}). 

Establishing the existence and nature of the unseen companion(s) can
provide important constraints on binary evolution, core explodability,
and the origin of compact binaries.  \azv\
{may offer}
an opportunity to directly observe the relationship between the binary companion's dynamical interaction and the disk ejection.
Since many classical OBe stars are massive, post-SN objects, it
suggests a likely link between OBe stars and massive,
{eccentric}
systems.
Further study of this fascinating object can more definitively confirm its status and exploit the opportunities it offers to learn about massive binary evolution and disk ejection.

\acknowledgements{
We benefited from useful discussions with many people, including Jon
Bjorkman, Paul Crowther, Julian Deman, Jim Fuller, Jay Gallagher,
Carol Jones, Max Moe, Megan Reiter,
{Steve Shore,}
and Drew Weisserman.
Many thanks to Juliette Becker for the use of her code, and to Traci Johnson, Mario Mateo, and the M2FS Team for help with observing runs.
We also thank the anonymous referees for
valuable comments that greatly improved this paper.
This work was supported by NSF grant
AST-1514838 to M.S.O. and by the University of Michigan.  N. Castro acknowledges funding from the Deutsche Forschungsgemeinschaft (DFG), CA 2551/1-1; M.R. is supported by EUH2020 OPTICON RadioNet Pilot grant No. 101004719; and R.H.D.T. is supported by NASA grant 80NSSC20K0515.
This research made use of Astropy, a community-developed core Python package for Astronomy
\citep{2013A&A...558A..33A}.  M.S.O. acknowledges MDRS, LLC, for pandemic hospitality.
}

\facilities{Magellan, OGLE, Gaia}
\clearpage

\bibliographystyle{aasjournal.bst}

\bibliography{77616_bib}

\begin{thebibliography}{}
\expandafter\ifx\csname natexlab\endcsname\relax\def\natexlab#1{#1}\fi

\bibitem[{{Alfonso-Garz{\'o}n} {et~al.}(2017){Alfonso-Garz{\'o}n}, {Fabregat},
  {Reig}, {Kajava}, {S{\'a}nchez-Fern{\'a}ndez}, {Townsend}, {Mas-Hesse},
  {Crawford}, {Kretschmar}, \& {Coe}}]{2017A&A...607A..52A}
{Alfonso-Garz{\'o}n}, J., {Fabregat}, J., {Reig}, P., {et~al.} 2017, \aap, 607,
  A52

\bibitem[{{Arzoumanian} {et~al.}(2002){Arzoumanian}, {Chernoff}, \&
  {Cordes}}]{arzoumanian:02}
{Arzoumanian}, Z., {Chernoff}, D.~F., \& {Cordes}, J.~M. 2002, \apj, 568, 289

\bibitem[{{Astropy Collaboration} {et~al.}(2013){Astropy Collaboration},
  {Robitaille}, {Tollerud}, {Greenfield}, {Droettboom}, {Bray}, {Aldcroft},
  {Davis}, {Ginsburg}, {Price-Whelan}, {Kerzendorf}, {Conley}, {Crighton},
  {Barbary}, {Muna}, {Ferguson}, {Grollier}, {Parikh}, {Nair}, {Unther},
  {Deil}, {Woillez}, {Conseil}, {Kramer}, {Turner}, {Singer}, {Fox}, {Weaver},
  {Zabalza}, {Edwards}, {Azalee Bostroem}, {Burke}, {Casey}, {Crawford},
  {Dencheva}, {Ely}, {Jenness}, {Labrie}, {Lim}, {Pierfederici}, {Pontzen},
  {Ptak}, {Refsdal}, {Servillat}, \& {Streicher}}]{2013A&A...558A..33A}
{Astropy Collaboration}, {Robitaille}, T.~P., {Tollerud}, E.~J., {et~al.} 2013,
  \aap, 558, A33

\bibitem[{{Atri} {et~al.}(2019){Atri}, {Miller-Jones}, {Bahramian}, {Plotkin},
  {Jonker}, {Nelemans}, {Maccarone}, {Sivakoff}, {Deller}, {Chaty}, {Torres},
  {Horiuchi}, {McCallum}, {Natusch}, {Phillips}, {Stevens}, \&
  {Weston}}]{atri:19}
{Atri}, P., {Miller-Jones}, J.~C.~A., {Bahramian}, A., {et~al.} 2019, \mnras,
  489, 3116

\bibitem[{{Azzopardi} {et~al.}(1975){Azzopardi}, {Vigneau}, \&
  {Macquet}}]{Azzopardi1975}
{Azzopardi}, M., {Vigneau}, J., \& {Macquet}, M. 1975, \aaps, 22, 285

\bibitem[{{Baade} {et~al.}(2016){Baade}, {Rivinius}, {Pigulski}, {Carciofi},
  {Martayan}, {Moffat}, {Wade}, {Weiss}, {Grunhut}, {Handler}, {Kuschnig},
  {Mehner}, {Pablo}, {Popowicz}, {Rucinski}, \& {Whittaker}}]{Baade2016}
{Baade}, D., {Rivinius}, T., {Pigulski}, A., {et~al.} 2016, \aap, 588, A56

\bibitem[{{Baade} {et~al.}(2018){Baade}, {Pigulski}, {Rivinius}, {Carciofi},
  {Panoglou}, {Ghoreyshi}, {Handler}, {Kuschnig}, {Moffat}, {Pablo},
  {Popowicz}, {Wade}, {Weiss}, \& {Zwintz}}]{Baade2018}
{Baade}, D., {Pigulski}, A., {Rivinius}, T., {et~al.} 2018, \aap, 610, A70

\bibitem[{{Becker} {et~al.}(2015){Becker}, {Johnson}, {Vanderburg}, \&
  {Morton}}]{Becker2015}
{Becker}, J.~C., {Johnson}, J.~A., {Vanderburg}, A., \& {Morton}, T.~D. 2015,
  \apjs, 217, 29

\bibitem[{{Belczynski} {et~al.}(2008){Belczynski}, {Kalogera}, {Rasio}, {Taam},
  {Zezas}, {Bulik}, {Maccarone}, \& {Ivanova}}]{belczynski:08}
{Belczynski}, K., {Kalogera}, V., {Rasio}, F.~A., {et~al.} 2008, \apjs, 174,
  223

\bibitem[{{Bernstein} {et~al.}(2003){Bernstein}, {Shectman}, {Gunnels},
  {Mochnacki}, \& {Athey}}]{2003SPIE.4841.1694B}
{Bernstein}, R., {Shectman}, S.~A., {Gunnels}, S.~M., {Mochnacki}, S., \&
  {Athey}, A.~E. 2003, in Instrument Design and Performance for
  Optical/Infrared Ground-based Telescopes. Edited by Iye, Masanori; Moorwood,
  Alan F. M. Proceedings of the SPIE, Volume 4841, pp. 1694-1704 (2003)., Vol.
  4841, 1694--1704

\bibitem[{{Bigelow} \& {Dressler}(2003)}]{2003SPIE.4841.1727B}
{Bigelow}, B.~C., \& {Dressler}, A.~M. 2003, in Instrument Design and
  Performance for Optical/Infrared Ground-based Telescopes. Edited by Iye,
  Masanori; Moorwood, Alan F. M. Proceedings of the SPIE, Volume 4841, pp.
  1727-1738 (2003)., Vol. 4841, 1727--1738

\bibitem[{{Blaauw}(1993)}]{blaauw:93}
{Blaauw}, A. 1993, in Astronomical Society of the Pacific Conference Series,
  Vol.~35, Massive Stars: Their Lives in the Interstellar Medium, ed. J.~P.
  {Cassinelli} \& E.~B. {Churchwell}, 207

\bibitem[{{Blanco-Cuaresma} {et~al.}(2014){Blanco-Cuaresma}, {Soubiran},
  {Heiter}, \& {Jofr{\'e}}}]{Blanco-Cuaresma2014}
{Blanco-Cuaresma}, S., {Soubiran}, C., {Heiter}, U., \& {Jofr{\'e}}, P. 2014,
  \aap, 569, A111

\bibitem[{{Bodensteiner} {et~al.}(2020){Bodensteiner}, {Shenar}, \&
  {Sana}}]{bodensteiner:20}
{Bodensteiner}, J., {Shenar}, T., \& {Sana}, H. 2020, \aap, 641, A42

\bibitem[{{Bonanos} {et~al.}(2010){Bonanos}, {Lennon}, {K{\"o}hlinger}, {van
  Loon}, {Massa}, {Sewilo}, {Evans}, {Panagia}, {Babler}, {Block}, {Bracker},
  {Engelbracht}, {Gordon}, {Hora}, {Indebetouw}, {Meade}, {Meixner}, {Misselt},
  {Robitaille}, {Shiao}, \& {Whitney}}]{2010AJ....140..416B}
{Bonanos}, A.~Z., {Lennon}, D.~J., {K{\"o}hlinger}, F., {et~al.} 2010, \aj,
  140, 416

\bibitem[{{Brandt} \& {Podsiadlowski}(1995)}]{brandt:95}
{Brandt}, N., \& {Podsiadlowski}, P. 1995, \mnras, 274, 461

\bibitem[{{Brott} {et~al.}(2011){Brott}, {de Mink}, {Cantiello}, {Langer}, {de
  Koter}, {Evans}, {Hunter}, {Trundle}, \& {Vink}}]{2011A&A...530A.115B}
{Brott}, I., {de Mink}, S.~E., {Cantiello}, M., {et~al.} 2011, \aap, 530, A115

\bibitem[{{Callister} {et~al.}(2020){Callister}, {Farr}, \&
  {Renzo}}]{callister:20}
{Callister}, T.~A., {Farr}, W.~M., \& {Renzo}, M. 2020, arXiv e-prints,
  arXiv:2011.09570

\bibitem[{{Cantiello} {et~al.}(2007){Cantiello}, {Yoon}, {Langer}, \&
  {Livio}}]{cantiello:07}
{Cantiello}, M., {Yoon}, S., {Langer}, N., \& {Livio}, M. 2007, \aap, 465, L29

\bibitem[{{Castro} {et~al.}(2018){Castro}, {Oey}, {Fossati}, \&
  {Langer}}]{Castro2018}
{Castro}, N., {Oey}, M.~S., {Fossati}, L., \& {Langer}, N. 2018, \apj, 868, 57

\bibitem[{{Che} {et~al.}(2012){Che}, {Monnier}, {Tycner}, {Kraus}, {Zavala},
  {Baron}, {Pedretti}, {ten Brummelaar}, {McAlister}, {Ridgway}, {Sturmann},
  {Sturmann}, \& {Turner}}]{Che2012}
{Che}, X., {Monnier}, J.~D., {Tycner}, C., {et~al.} 2012, \apj, 757, 29

\bibitem[{{Conti} \& {Leep}(1974)}]{1974ApJ...193..113C}
{Conti}, P.~S., \& {Leep}, E.~M. 1974, \apj, 193, 113

\bibitem[{{Couch} {et~al.}(2020){Couch}, {Warren}, \& {O'Connor}}]{couch:20}
{Couch}, S.~M., {Warren}, M.~L., \& {O'Connor}, E.~P. 2020, \apj, 890, 127

\bibitem[{{Dallas} \& {Oey}(2022)}]{Dallas2022}
{Dallas}, M.~M., \& {Oey}, M.~S. 2022, \apj, in press

\bibitem[{{de Wit} {et~al.}(2006){de Wit}, {Lamers}, {Marquette}, \&
  {Beaulieu}}]{2006A&A...456.1027D}
{de Wit}, W.~J., {Lamers}, H.~J.~G.~L.~M., {Marquette}, J.~B., \& {Beaulieu},
  J.~P. 2006, \aap, 456, 1027

\bibitem[{{Dorigo Jones} {et~al.}(2020){Dorigo Jones}, {Oey}, {Paggeot},
  {Castro}, \& {Moe}}]{DorigoJones2020}
{Dorigo Jones}, J., {Oey}, M.~S., {Paggeot}, K., {Castro}, N., \& {Moe}, M.
  2020, \apj, 903, 43

\bibitem[{{Dray} {et~al.}(2005){Dray}, {Dale}, {Beer}, {Napiwotzki}, \&
  {King}}]{dray:05}
{Dray}, L.~M., {Dale}, J.~E., {Beer}, M.~E., {Napiwotzki}, R., \& {King}, A.~R.
  2005, \mnras, 364, 59

\bibitem[{{Fitzpatrick} \& {Massa}(2007)}]{2007ApJ...663..320F}
{Fitzpatrick}, E.~L., \& {Massa}, D. 2007, \apj, 663, 320

\bibitem[{{Gaia Collaboration} {et~al.}(2021){Gaia Collaboration}, {Brown},
  {Vallenari}, {Prusti}, {de Bruijne}, {Babusiaux}, {Biermann}, {Creevey},
  {Evans}, {Eyer}, {Hutton}, {Jansen}, {Jordi}, {Klioner}, {Lammers},
  {Lindegren}, {Luri}, {Mignard}, {Panem}, {Pourbaix}, {Randich}, {Sartoretti},
  {Soubiran}, {Walton}, {Arenou}, {Bailer-Jones}, {Bastian}, {Cropper},
  {Drimmel}, {Katz}, {Lattanzi}, {van Leeuwen}, {Bakker}, {Cacciari},
  {Casta{\~n}eda}, {De Angeli}, {Ducourant}, {Fabricius}, {Fouesneau},
  {Fr{\'e}mat}, {Guerra}, {Guerrier}, {Guiraud}, {Jean-Antoine Piccolo},
  {Masana}, {Messineo}, {Mowlavi}, {Nicolas}, {Nienartowicz}, {Pailler},
  {Panuzzo}, {Riclet}, {Roux}, {Seabroke}, {Sordo}, {Tanga}, {Th{\'e}venin},
  {Gracia-Abril}, {Portell}, {Teyssier}, {Altmann}, {Andrae}, {Bellas-Velidis},
  {Benson}, {Berthier}, {Blomme}, {Brugaletta}, {Burgess}, {Busso}, {Carry},
  {Cellino}, {Cheek}, {Clementini}, {Damerdji}, {Davidson}, {Delchambre},
  {Dell'Oro}, {Fern{\'a}ndez-Hern{\'a}ndez}, {Galluccio}, {Garc{\'\i}a-Lario},
  {Garcia-Reinaldos}, {Gonz{\'a}lez-N{\'u}{\~n}ez}, {Gosset}, {Haigron},
  {Halbwachs}, {Hambly}, {Harrison}, {Hatzidimitriou}, {Heiter},
  {Hern{\'a}ndez}, {Hestroffer}, {Hodgkin}, {Holl}, {Jan{\ss}en}, {Jevardat de
  Fombelle}, {Jordan}, {Krone-Martins}, {Lanzafame}, {L{\"o}ffler}, {Lorca},
  {Manteiga}, {Marchal}, {Marrese}, {Moitinho}, {Mora}, {Muinonen}, {Osborne},
  {Pancino}, {Pauwels}, {Petit}, {Recio-Blanco}, {Richards}, {Riello},
  {Rimoldini}, {Robin}, {Roegiers}, {Rybizki}, {Sarro}, {Siopis}, {Smith},
  {Sozzetti}, {Ulla}, {Utrilla}, {van Leeuwen}, {van Reeven}, {Abbas}, {Abreu
  Aramburu}, {Accart}, {Aerts}, {Aguado}, {Ajaj}, {Altavilla}, {{\'A}lvarez},
  {{\'A}lvarez Cid-Fuentes}, {Alves}, {Anderson}, {Anglada Varela}, {Antoja},
  {Audard}, {Baines}, {Baker}, {Balaguer-N{\'u}{\~n}ez}, {Balbinot}, {Balog},
  {Barache}, {Barbato}, {Barros}, {Barstow}, {Bartolom{\'e}}, {Bassilana},
  {Bauchet}, {Baudesson-Stella}, {Becciani}, {Bellazzini}, {Bernet}, {Bertone},
  {Bianchi}, {Blanco-Cuaresma}, {Boch}, {Bombrun}, {Bossini}, {Bouquillon},
  {Bragaglia}, {Bramante}, {Breedt}, {Bressan}, {Brouillet}, {Bucciarelli},
  {Burlacu}, {Busonero}, {Butkevich}, {Buzzi}, {Caffau}, {Cancelliere},
  {C{\'a}novas}, {Cantat-Gaudin}, {Carballo}, {Carlucci}, {Carnerero},
  {Carrasco}, {Casamiquela}, {Castellani}, {Castro-Ginard}, {Castro Sampol},
  {Chaoul}, {Charlot}, {Chemin}, {Chiavassa}, {Cioni}, {Comoretto}, {Cooper},
  {Cornez}, {Cowell}, {Crifo}, {Crosta}, {Crowley}, {Dafonte}, {Dapergolas},
  {David}, {David}, {de Laverny}, {De Luise}, {De March}, {De Ridder}, {de
  Souza}, {de Teodoro}, {de Torres}, {del Peloso}, {del Pozo}, {Delbo},
  {Delgado}, {Delgado}, {Delisle}, {Di Matteo}, {Diakite}, {Diener},
  {Distefano}, {Dolding}, {Eappachen}, {Edvardsson}, {Enke}, {Esquej}, {Fabre},
  {Fabrizio}, {Faigler}, {Fedorets}, {Fernique}, {Fienga}, {Figueras},
  {Fouron}, {Fragkoudi}, {Fraile}, {Franke}, {Gai}, {Garabato},
  {Garcia-Gutierrez}, {Garc{\'\i}a-Torres}, {Garofalo}, {Gavras}, {Gerlach},
  {Geyer}, {Giacobbe}, {Gilmore}, {Girona}, {Giuffrida}, {Gomel}, {Gomez},
  {Gonzalez-Santamaria}, {Gonz{\'a}lez-Vidal}, {Granvik},
  {Guti{\'e}rrez-S{\'a}nchez}, {Guy}, {Hauser}, {Haywood}, {Helmi}, {Hidalgo},
  {Hilger}, {H{\l}adczuk}, {Hobbs}, {Holland}, {Huckle}, {Jasniewicz},
  {Jonker}, {Juaristi Campillo}, {Julbe}, {Karbevska}, {Kervella}, {Khanna},
  {Kochoska}, {Kontizas}, {Kordopatis}, {Korn}, {Kostrzewa-Rutkowska},
  {Kruszy{\'n}ska}, {Lambert}, {Lanza}, {Lasne}, {Le Campion}, {Le Fustec},
  {Lebreton}, {Lebzelter}, {Leccia}, {Leclerc}, {Lecoeur-Taibi}, {Liao},
  {Licata}, {Lindstr{\o}m}, {Lister}, {Livanou}, {Lobel}, {Madrero Pardo},
  {Managau}, {Mann}, {Marchant}, {Marconi}, {Marcos Santos}, {Marinoni},
  {Marocco}, {Marshall}, {Martin Polo}, {Mart{\'\i}n-Fleitas}, {Masip},
  {Massari}, {Mastrobuono-Battisti}, {Mazeh}, {McMillan}, {Messina},
  {Michalik}, {Millar}, {Mints}, {Molina}, {Molinaro}, {Moln{\'a}r},
  {Montegriffo}, {Mor}, {Morbidelli}, {Morel}, {Morris}, {Mulone}, {Munoz},
  {Muraveva}, {Murphy}, {Musella}, {Noval}, {Ord{\'e}novic}, {Orr{\`u}},
  {Osinde}, {Pagani}, {Pagano}, {Palaversa}, {Palicio}, {Panahi}, {Pawlak},
  {Pe{\~n}alosa Esteller}, {Penttil{\"a}}, {Piersimoni}, {Pineau}, {Plachy},
  {Plum}, {Poggio}, {Poretti}, {Poujoulet}, {Pr{\v{s}}a}, {Pulone}, {Racero},
  {Ragaini}, {Rainer}, {Raiteri}, {Rambaux}, {Ramos}, {Ramos-Lerate}, {Re
  Fiorentin}, {Regibo}, {Reyl{\'e}}, {Ripepi}, {Riva}, {Rixon}, {Robichon},
  {Robin}, {Roelens}, {Rohrbasser}, {Romero-G{\'o}mez}, {Rowell}, {Royer},
  {Rybicki}, {Sadowski}, {Sagrist{\`a} Sell{\'e}s}, {Sahlmann}, {Salgado},
  {Salguero}, {Samaras}, {Sanchez Gimenez}, {Sanna}, {Santove{\~n}a},
  {Sarasso}, {Schultheis}, {Sciacca}, {Segol}, {Segovia}, {S{\'e}gransan},
  {Semeux}, {Shahaf}, {Siddiqui}, {Siebert}, {Siltala}, {Slezak}, {Smart},
  {Solano}, {Solitro}, {Souami}, {Souchay}, {Spagna}, {Spoto}, {Steele},
  {Steidelm{\"u}ller}, {Stephenson}, {S{\"u}veges}, {Szabados}, {Szegedi-Elek},
  {Taris}, {Tauran}, {Taylor}, {Teixeira}, {Thuillot}, {Tonello}, {Torra},
  {Torra}, {Turon}, {Unger}, {Vaillant}, {van Dillen}, {Vanel}, {Vecchiato},
  {Viala}, {Vicente}, {Voutsinas}, {Weiler}, {Wevers}, {Wyrzykowski}, {Yoldas},
  {Yvard}, {Zhao}, {Zorec}, {Zucker}, {Zurbach}, \& {Zwitter}}]{Gaia2021}
{Gaia Collaboration}, {Brown}, A.~G.~A., {Vallenari}, A., {et~al.} 2021, \aap,
  649, A1

\bibitem[{{Ghoreyshi} {et~al.}(2018){Ghoreyshi}, {Carciofi}, {R{\'\i}mulo},
  {Vieira}, {Faes}, {Baade}, {Bjorkman}, {Otero}, \&
  {Rivinius}}]{Ghoreyshi2018}
{Ghoreyshi}, M.~R., {Carciofi}, A.~C., {R{\'\i}mulo}, L.~R., {et~al.} 2018,
  \mnras, 479, 2214

\bibitem[{{Golden-Marx} {et~al.}(2016){Golden-Marx}, {Oey}, {Lamb}, {Graus}, \&
  {White}}]{2016ApJ...819...55G}
{Golden-Marx}, J.~B., {Oey}, M.~S., {Lamb}, J.~B., {Graus}, A.~S., \& {White},
  A.~S. 2016, \apj, 819, 55

\bibitem[{{G{\"o}tberg} {et~al.}(2017){G{\"o}tberg}, {de Mink}, \&
  {Groh}}]{gotberg:17}
{G{\"o}tberg}, Y., {de Mink}, S.~E., \& {Groh}, J.~H. 2017, \aap, 608, A11

\bibitem[{{Graczyk} {et~al.}(2014){Graczyk}, {Pietrzy{\'n}ski}, {Thompson},
  {Gieren}, {Pilecki}, {Konorski}, {Udalski}, {Soszy{\'n}ski}, {Villanova},
  {G{\'o}rski}, {Suchomska}, {Karczmarek}, {Kudritzki}, {Bresolin}, \&
  {Gallenne}}]{2014ApJ...780...59G}
{Graczyk}, D., {Pietrzy{\'n}ski}, G., {Thompson}, I.~B., {et~al.} 2014, \apj,
  780, 59

\bibitem[{{Hainich} {et~al.}(2019){Hainich}, {Ramachandran}, {Shenar},
  {Sander}, {Todt}, {Gruner}, {Oskinova}, \& {Hamann}}]{Hainich2019}
{Hainich}, R., {Ramachandran}, V., {Shenar}, T., {et~al.} 2019, \aap, 621, A85

\bibitem[{{Hartmann} {et~al.}(2016){Hartmann}, {Herczeg}, \&
  {Calvet}}]{Hartmann2016}
{Hartmann}, L., {Herczeg}, G., \& {Calvet}, N. 2016, \araa, 54, 135

\bibitem[{{Henize}(1956)}]{Henize1956}
{Henize}, K.~G. 1956, \apjs, 2, 315

\bibitem[{{Hirai} {et~al.}(2021){Hirai}, {Podsiadlowski}, {Owocki},
  {Schneider}, \& {Smith}}]{Hirai2021}
{Hirai}, R., {Podsiadlowski}, P., {Owocki}, S.~P., {Schneider}, F. R.~N., \&
  {Smith}, N. 2021, \mnras, 503, 4276

\bibitem[{{Janka}(2013)}]{janka:13}
{Janka}, H.-T. 2013, \mnras, 434, 1355

\bibitem[{{Janka}(2017)}]{Janka2017}
---. 2017, \apj, 837, 84

\bibitem[{{Jayasinghe} {et~al.}(2021){Jayasinghe}, {Kochanek}, {Strader},
  {Stanek}, {Vallely}, {Thompson}, {Hinkle}, {Shappee}, {Dupree}, {Auchettl},
  {Chomiuk}, {Aydi}, {Dage}, {Hughes}, {Shishkovsky}, {Sokolovsky}, {Swihart},
  {Voggel}, \& {Thompson}}]{Jayasinghe2021}
{Jayasinghe}, T., {Kochanek}, C.~S., {Strader}, J., {et~al.} 2021, \mnras, 506,
  4083

\bibitem[{{Katahira} {et~al.}(1996){Katahira}, {Hirata}, {Ito}, {Katoh},
  {Ballereau}, \& {Chauville}}]{Katahira1996}
{Katahira}, J.-I., {Hirata}, R., {Ito}, M., {et~al.} 1996, \pasj, 48, 317

\bibitem[{{Kelson}(2003)}]{2003PASP..115..688K}
{Kelson}, D.~D. 2003, Publications of the Astronomical Society of the Pacific,
  115, 688

\bibitem[{{Kelson} {et~al.}(2000){Kelson}, {Illingworth}, {van Dokkum}, \&
  {Franx}}]{2000ApJ...531..159K}
{Kelson}, D.~D., {Illingworth}, G.~D., {van Dokkum}, P.~G., \& {Franx}, M.
  2000, \apj, 531, 159

\bibitem[{{Klencki} {et~al.}(2022){Klencki}, {Istrate}, {Nelemans}, \&
  {Pols}}]{Klencki2022}
{Klencki}, J., {Istrate}, A., {Nelemans}, G., \& {Pols}, O. 2022, \aap, 662,
  A56

\bibitem[{{Labadie-Bartz} {et~al.}(2017){Labadie-Bartz}, {Pepper}, {McSwain},
  {Bjorkman}, {Bjorkman}, {Lund}, {Rodriguez}, {Stassun}, {Stevens}, {James},
  {Kuhn}, {Siverd}, \& {Beatty}}]{Labadie2017}
{Labadie-Bartz}, J., {Pepper}, J., {McSwain}, M.~V., {et~al.} 2017, \aj, 153,
  252

\bibitem[{{Lamb} {et~al.}(2016){Lamb}, {Oey}, {Segura-Cox}, {Graus}, {Kiminki},
  {Golden-Marx}, \& {Parker}}]{2016ApJ...817..113L}
{Lamb}, J.~B., {Oey}, M.~S., {Segura-Cox}, D.~M., {et~al.} 2016, \apj, 817, 113

\bibitem[{{Mandel}(2016)}]{mandel:16}
{Mandel}, I. 2016, \mnras, 456, 578

\bibitem[{{Marr} {et~al.}(2022){Marr}, {Jones}, {Tycner}, {Carciofi}, \&
  {Silva}}]{Marr2022}
{Marr}, K.~C., {Jones}, C.~E., {Tycner}, C., {Carciofi}, A.~C., \& {Silva},
  A.~C.~F. 2022, \apj, 928, 145

\bibitem[{{Martins} \& {Palacios}(2021)}]{Martins2021}
{Martins}, F., \& {Palacios}, A. 2021, \aap, 645, A67

\bibitem[{{Martins} {et~al.}(2005){Martins}, {Schaerer}, \&
  {Hillier}}]{2005A&A...436.1049M}
{Martins}, F., {Schaerer}, D., \& {Hillier}, D.~J. 2005, \aap, 436, 1049

\bibitem[{{Massey}(2002)}]{2002ApJS..141...81M}
{Massey}, P. 2002, The Astrophysical Journal Supplement Series, 141, 81

\bibitem[{{Mateo} {et~al.}(2012){Mateo}, {Bailey}, {Crane}, {Shectman},
  {Thompson}, {Roederer}, {Bigelow}, \& {Gunnels}}]{2012SPIE.8446E..4YM}
{Mateo}, M., {Bailey}, J.~I., {Crane}, J., {et~al.} 2012, in Proceedings of the
  SPIE, Volume 8446, id. 84464Y 19 pp. (2012)., Vol. 8446, 84464Y

\bibitem[{{Moe} \& {Di Stefano}(2017)}]{Moe2017}
{Moe}, M., \& {Di Stefano}, R. 2017, \apjs, 230, 15

\bibitem[{{Natta} \& {Whitney}(2000)}]{Natta2000}
{Natta}, A., \& {Whitney}, B.~A. 2000, \aap, 364, 633

\bibitem[{{Nemravov{\'a}} {et~al.}(2010){Nemravov{\'a}}, {Harmanec},
  {Kub{\'a}t}, {Koubsk{\'y}}, {Iliev}, {Yang}, {Ribeiro}, {{\v{S}}lechta},
  {Kotkov{\'a}}, {Wolf}, \& {{\v{S}}koda}}]{Nemravova2010}
{Nemravov{\'a}}, J., {Harmanec}, P., {Kub{\'a}t}, J., {et~al.} 2010, \aap, 516,
  A80

\bibitem[{{O'Connor} \& {Ott}(2011)}]{oconnor2011}
{O'Connor}, E., \& {Ott}, C.~D. 2011, \apj, 730, 70

\bibitem[{{Okazaki}(1991)}]{1991PASJ...43...75O}
{Okazaki}, A.~T. 1991, Publications of the Astronomical Society of Japan, 43,
  75

\bibitem[{{Pablo} {et~al.}(2017){Pablo}, {Richardson}, {Fuller}, {Rowe},
  {Moffat}, {Kuschnig}, {Popowicz}, {Handler}, {Neiner}, {Pigulski}, {Wade},
  {Weiss}, {Buysschaert}, {Ramiaramanantsoa}, {Bratcher}, {Gerhartz}, {Greco},
  {Hardegree-Ullman}, {Lembryk}, \& {Oswald}}]{Pablo2017}
{Pablo}, H., {Richardson}, N.~D., {Fuller}, J., {et~al.} 2017, \mnras, 467,
  2494

\bibitem[{{Packet}(1981)}]{packet:81}
{Packet}, W. 1981, \aap, 102, 17

\bibitem[{{Pflamm-Altenburg} \& {Kroupa}(2010)}]{Pflamm2010}
{Pflamm-Altenburg}, J., \& {Kroupa}, P. 2010, \mnras, 404, 1564

\bibitem[{{Podsiadlowski} {et~al.}(2004){Podsiadlowski}, {Langer},
  {Poelarends}, {Rappaport}, {Heger}, \& {Pfahl}}]{podsiadlowski:04}
{Podsiadlowski}, P., {Langer}, N., {Poelarends}, A.~J.~T., {et~al.} 2004, \apj,
  612, 1044

\bibitem[{{Poeckert}(1982)}]{1982IAUS...98..453P}
{Poeckert}, R. 1982, in Be Stars, Vol.~98, 453--477

\bibitem[{{Poleski} {et~al.}(2012){Poleski}, {Soszy{\'n}ski}, {Udalski},
  {Szyma{\'n}ski}, {Kubiak}, {Pietrzy{\'n}ski}, {Wyrzykowski}, \&
  {Ulaczyk}}]{2012AcA....62....1P}
{Poleski}, R., {Soszy{\'n}ski}, I., {Udalski}, A., {et~al.} 2012, \actaa, 62, 1

\bibitem[{{Pols} {et~al.}(1991){Pols}, {Cote}, {Waters}, \& {Heise}}]{pols:91}
{Pols}, O.~R., {Cote}, J., {Waters}, L.~B.~F.~M., \& {Heise}, J. 1991, \aap,
  241, 419

\bibitem[{{Puls} {et~al.}(2005){Puls}, {Urbaneja}, {Venero}, {Repolust},
  {Springmann}, {Jokuthy}, \& {Mokiem}}]{2005A&A...435..669P}
{Puls}, J., {Urbaneja}, M.~A., {Venero}, R., {et~al.} 2005, \aap, 435, 669

\bibitem[{{Renzo} \& {G{\"o}tberg}(2021)}]{renzo:21zeta}
{Renzo}, M., \& {G{\"o}tberg}, Y. 2021, arXiv e-prints, arXiv:2107.10933

\bibitem[{{Renzo} {et~al.}(2019){Renzo}, {Zapartas}, {de Mink}, {G{\"o}tberg},
  {Justham}, {Farmer}, {Izzard}, {Toonen}, \& {Sana}}]{renzo:19walk}
{Renzo}, M., {Zapartas}, E., {de Mink}, S.~E., {et~al.} 2019, \aap, 624, A66

\bibitem[{{Repetto} {et~al.}(2017){Repetto}, {Igoshev}, \&
  {Nelemans}}]{repetto:17}
{Repetto}, S., {Igoshev}, A.~P., \& {Nelemans}, G. 2017, \mnras, 467, 298

\bibitem[{{Richardson} {et~al.}(2021){Richardson}, {Thizy}, {Bjorkman},
  {Carciofi}, {Rubio}, {Thomas}, {Bjorkman}, {Labadie-Bartz}, {Genaro},
  {Wisniewski}, {Wang}, {Gies}, {Chojnowski}, {Daly}, {Edwards}, {Fowler},
  {Gullingsrud}, {Habel}, {James}, {Kehoe}, {Kuchta}, {Lane}, {Miroshnichenko},
  {Mishra}, {Pablo}, {Peploski}, {Pepper}, {Rodriguez}, {Siverd}, {Stassun},
  {Stevens}, {Trucks}, {Windsor}, {Wood}, {Bertrand}, {Broussat}, {Bryssinck},
  {Buil}, {Charbonnel}, {de Bruin}, {Daglen}, {Desnoux}, {Dull}, {Garde},
  {Graham}, {Gurney}, {Halsey}, {Fosanelli}, {Guarro Fl{\'o}}, {Houpert},
  {James}, {Kreider}, {Leadbeater}, {Lester}, {Li}, {Maetz}, {Stiewing},
  {Somogyi}, {Terry}, {Ubaud}, \& {Waldschlaeger}}]{richardson:21}
{Richardson}, N.~D., {Thizy}, O., {Bjorkman}, J.~E., {et~al.} 2021, arXiv
  e-prints, arXiv:2109.11026

\bibitem[{{Rivero Gonz{\'a}lez} {et~al.}(2012){Rivero Gonz{\'a}lez}, {Puls},
  {Massey}, \& {Najarro}}]{2012A&A...543A..95R}
{Rivero Gonz{\'a}lez}, J.~G., {Puls}, J., {Massey}, P., \& {Najarro}, F. 2012,
  \aap, 543, A95

\bibitem[{{Rivinius} {et~al.}(2013){Rivinius}, {Carciofi}, \&
  {Martayan}}]{2013A&ARv..21...69R}
{Rivinius}, T., {Carciofi}, A.~C., \& {Martayan}, C. 2013, Astronomy and
  Astrophysics Review, 21, 69

\bibitem[{{Santolaya-Rey} {et~al.}(1997){Santolaya-Rey}, {Puls}, \&
  {Herrero}}]{1997A&A...323..488S}
{Santolaya-Rey}, A.~E., {Puls}, J., \& {Herrero}, A. 1997, \aap, 323, 488

\bibitem[{{Schootemeijer} {et~al.}(2018){Schootemeijer}, {G{\"o}tberg}, {de
  Mink}, {Gies}, \& {Zapartas}}]{schootemeijer:2018}
{Schootemeijer}, A., {G{\"o}tberg}, Y., {de Mink}, S.~E., {Gies}, D., \&
  {Zapartas}, E. 2018, \aap, 615, A30

\bibitem[{{Sigut} {et~al.}(2009){Sigut}, {McGill}, \& {Jones}}]{Sigut2009}
{Sigut}, T.~A.~A., {McGill}, M.~A., \& {Jones}, C.~E. 2009, \apj, 699, 1973

\bibitem[{{Sim{\'o}n-D{\'\i}az} \& {Herrero}(2007)}]{2007A&A...468.1063S}
{Sim{\'o}n-D{\'\i}az}, S., \& {Herrero}, A. 2007, \aap, 468, 1063

\bibitem[{{Sim{\'o}n-D{\'\i}az} \& {Herrero}(2014)}]{2014A&A...562A.135S}
---. 2014, \aap, 562, A135

\bibitem[{{Sim{\'o}n-D{\'\i}az} {et~al.}(2015){Sim{\'o}n-D{\'\i}az},
  {Caballero}, {Lorenzo}, {Ma{\'\i}z Apell{\'a}niz}, {Schneider}, {Negueruela},
  {Barb{\'a}}, {Dorda}, {Marco}, {Montes}, {Pellerin}, {Sanchez-Bermudez},
  {S{\'o}dor}, \& {Sota}}]{SimonDiaz2015}
{Sim{\'o}n-D{\'\i}az}, S., {Caballero}, J.~A., {Lorenzo}, J., {et~al.} 2015,
  \apj, 799, 169

\bibitem[{{Skrutskie} {et~al.}(2006){Skrutskie}, {Cutri}, {Stiening},
  {Weinberg}, {Schneider}, {Carpenter}, {Beichman}, {Capps}, {Chester},
  {Elias}, {Huchra}, {Liebert}, {Lonsdale}, {Monet}, {Price}, {Seitzer},
  {Jarrett}, {Kirkpatrick}, {Gizis}, {Howard}, {Evans}, {Fowler}, {Fullmer},
  {Hurt}, {Light}, {Kopan}, {Marsh}, {McCallon}, {Tam}, {Van Dyk}, \&
  {Wheelock}}]{2006AJ....131.1163S}
{Skrutskie}, M.~F., {Cutri}, R.~M., {Stiening}, R., {et~al.} 2006, \aj, 131,
  1163

\bibitem[{{Smith} {et~al.}(2005){Smith}, {Points}, {Chu}, {Winkler},
  {Aguilera}, {Leiton}, \& {MCELS Team}}]{Smith2005}
{Smith}, R.~C., {Points}, S.~D., {Chu}, Y.~H., {et~al.} 2005, in American
  Astronomical Society Meeting Abstracts, Vol. 207, American Astronomical
  Society Meeting Abstracts, 25.07

\bibitem[{{Sota} {et~al.}(2011){Sota}, {Ma{\'\i}z Apell{\'a}niz}, {Walborn},
  {Alfaro}, {Barb{\'a}}, {Morrell}, {Gamen}, \& {Arias}}]{Sota2011}
{Sota}, A., {Ma{\'\i}z Apell{\'a}niz}, J., {Walborn}, N.~R., {et~al.} 2011,
  \apjs, 193, 24

\bibitem[{{Suffak} {et~al.}(2022){Suffak}, {Jones}, \& {Carciofi}}]{Suffak2022}
{Suffak}, M., {Jones}, C.~E., \& {Carciofi}, A.~C. 2022, \mnras, 509, 931

\bibitem[{{Suffak} {et~al.}(2020){Suffak}, {Jones}, {Tycner}, {Henry},
  {Carciofi}, {Mota}, \& {Rubio}}]{Suffak2020}
{Suffak}, M.~W., {Jones}, C.~E., {Tycner}, C., {et~al.} 2020, \apj, 890, 86

\bibitem[{{Sukhbold} {et~al.}(2016){Sukhbold}, {Ertl}, {Woosley}, {Brown}, \&
  {Janka}}]{sukhbold:16}
{Sukhbold}, T., {Ertl}, T., {Woosley}, S.~E., {Brown}, J.~M., \& {Janka}, H.-T.
  2016, \apj, 821, 38

\bibitem[{{Sun} {et~al.}(2023){Sun}, {Townsend}, \& {Guo}}]{Sun2023}
{Sun}, M., {Townsend}, R.~H.~D., \& {Guo}, Z. 2023, arXiv e-prints,
  arXiv:2301.06599

\bibitem[{{Tango} {et~al.}(2009){Tango}, {Davis}, {Jacob}, {Mendez}, {North},
  {O'Byrne}, {Seneta}, \& {Tuthill}}]{Tango2009}
{Tango}, W.~J., {Davis}, J., {Jacob}, A.~P., {et~al.} 2009, \mnras, 396, 842

\bibitem[{{Tauris} \& {Takens}(1998)}]{tauris:98}
{Tauris}, T.~M., \& {Takens}, R.~J. 1998, \aap, 330, 1047

\bibitem[{{Toonen} {et~al.}(2020){Toonen}, {Portegies Zwart}, {Hamers}, \&
  {Bandopadhyay}}]{toonen:20}
{Toonen}, S., {Portegies Zwart}, S., {Hamers}, A.~S., \& {Bandopadhyay}, D.
  2020, \aap, 640, A16

\bibitem[{{Tycner} {et~al.}(2011){Tycner}, {Ames}, {Zavala}, {Hummel},
  {Benson}, \& {Hutter}}]{Tycner2011}
{Tycner}, C., {Ames}, A., {Zavala}, R.~T., {et~al.} 2011, \apjl, 729, L5

\bibitem[{{Udalski} {et~al.}(2008){Udalski}, {Szymanski}, {Soszynski}, \&
  {Poleski}}]{2008AcA....58...69U}
{Udalski}, A., {Szymanski}, M.~K., {Soszynski}, I., \& {Poleski}, R. 2008,
  \actaa, 58, 69

\bibitem[{{Udalski} {et~al.}(2015){Udalski}, {Szyma{\'n}ski}, \&
  {Szyma{\'n}ski}}]{Udalski2015}
{Udalski}, A., {Szyma{\'n}ski}, M.~K., \& {Szyma{\'n}ski}, G. 2015, \actaa, 65,
  1

\bibitem[{{van den Heuvel}(1969)}]{vandenHeuvel1969}
{van den Heuvel}, E.~P.~J. 1969, \aj, 74, 1095

\bibitem[{{van der Meij} {et~al.}(2021){van der Meij}, {Guo}, {Kaper}, \&
  {Renzo}}]{vandermeij:21}
{van der Meij}, V., {Guo}, D., {Kaper}, L., \& {Renzo}, M. 2021, arXiv
  e-prints, arXiv:2108.12918

\bibitem[{{Vargas-Salazar} {et~al.}(2020){Vargas-Salazar}, {Oey}, {Barnes},
  {Chen}, {Castro}, {Kratter}, \& {Faerber}}]{Vargas2020}
{Vargas-Salazar}, I., {Oey}, M.~S., {Barnes}, J.~R., {et~al.} 2020, \apj, 903,
  42

\bibitem[{{Verbunt} {et~al.}(2017){Verbunt}, {Igoshev}, \&
  {Cator}}]{verbunt:17b}
{Verbunt}, F., {Igoshev}, A., \& {Cator}, E. 2017, \aap, 608, A57

\bibitem[{{Vinciguerra} {et~al.}(2020){Vinciguerra}, {Neijssel},
  {Vigna-G{\'o}mez}, {Mandel}, {Podsiadlowski}, {Maccarone}, {Nicholl},
  {Kingdon}, {Perry}, \& {Salemi}}]{vinciguerra:20}
{Vinciguerra}, S., {Neijssel}, C.~J., {Vigna-G{\'o}mez}, A., {et~al.} 2020,
  \mnras, 498, 4705

\bibitem[{{Walker} {et~al.}(2015){Walker}, {Mateo}, {Olszewski}, {Bailey},
  {Koposov}, {Belokurov}, \& {Evans}}]{2015ApJ...808..108W}
{Walker}, M.~G., {Mateo}, M., {Olszewski}, E.~W., {et~al.} 2015, \apj, 808, 108

\bibitem[{{Wang} {et~al.}(2021){Wang}, {Gies}, {Peters}, {G{\"o}tberg},
  {Chojnowski}, {Lester}, \& {Howell}}]{Wang2021}
{Wang}, L., {Gies}, D.~R., {Peters}, G.~J., {et~al.} 2021, \aj, 161, 248

\bibitem[{{Zapartas} {et~al.}(2021){Zapartas}, {Renzo}, {Fragos}, {Dotter},
  {Andrews}, {Bavera}, {Coughlin}, {Misra}, {Kovlakas}, {Rom{\'a}n-Garza},
  {Serra}, {Qin}, {Rocha}, \& {Tran}}]{zapartas:21}
{Zapartas}, E., {Renzo}, M., {Fragos}, T., {et~al.} 2021, arXiv e-prints,
  arXiv:2106.05228

\bibitem[{{Zechmeister} \& {K{\"u}rster}(2009)}]{Zechmeister2009}
{Zechmeister}, M., \& {K{\"u}rster}, M. 2009, \aap, 496, 577

\bibitem[{{Zhang} {et~al.}(2008){Zhang}, {Woosley}, \& {Heger}}]{zhang2008}
{Zhang}, W., {Woosley}, S.~E., \& {Heger}, A. 2008, \apj, 679, 639

\end{thebibliography}

\appendix
\section{Generalized Lomb-Scargle periodograms}

Figure~\ref{fig:GLSfits} shows the individual generalized Lomb-Scargle periodograms \citep{Zechmeister2009} and ancillary information for the six, roughly contiguous, OGLE datasets during  $\sim$ 2010 -- 2016 (Section~\ref{Sect:Pulsation}). 

\begin{figure*}[h]
\vspace*{-0.3in}
  \plotone{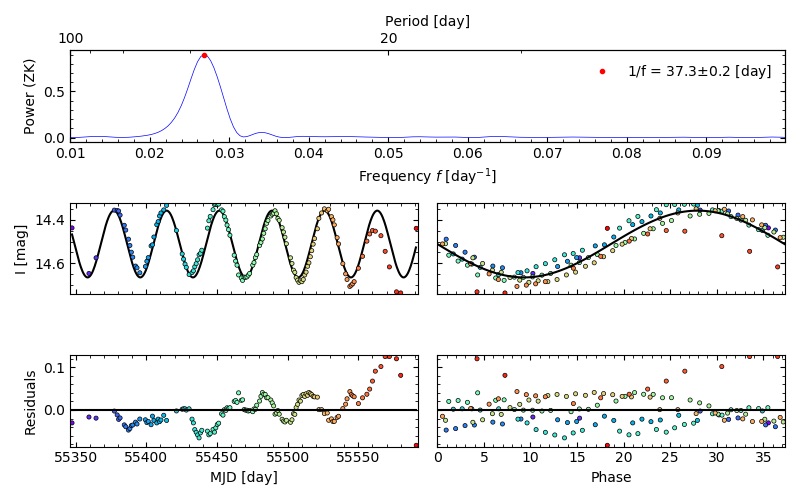}
    \plotone{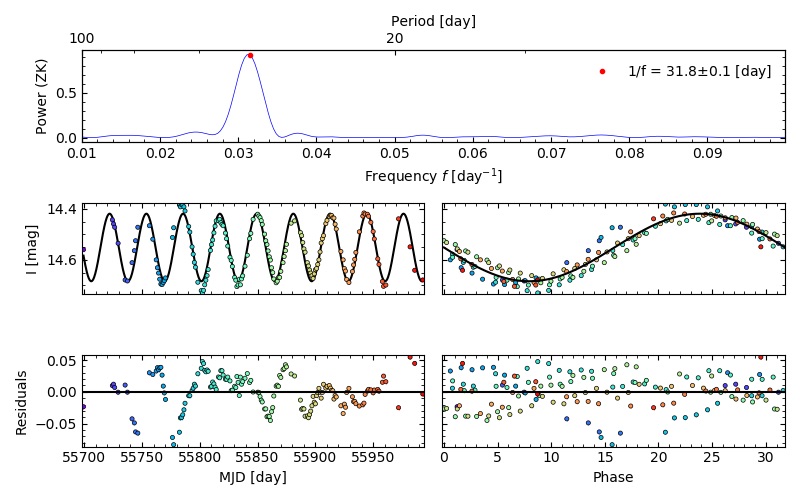}
% \vspace*{-1in}
	\caption{Top panels show the generalized Lomb-Scargle periodogram for light curves shown in the middle-left panels.  The fitted light curves are shown in the middle-right panels, with each cycle superposed according to color from the middle-left panel.  Residuals are shown in the bottom panels, as a function of MJD and phase, as shown.  The middle and bottom panels have the same $x$-axes.  The fitted period is shown in the top panel as the inverse of the frequency $f$.  The observation time of spectroscopic epoch~B is shown by the vertical dashed line in the plots for the fifth dataset.
	\label{fig:GLSfits}}
\end{figure*}

\begin{figure*}[h]
\figurenum{13}
\vspace*{-0.3in}
    \plotone{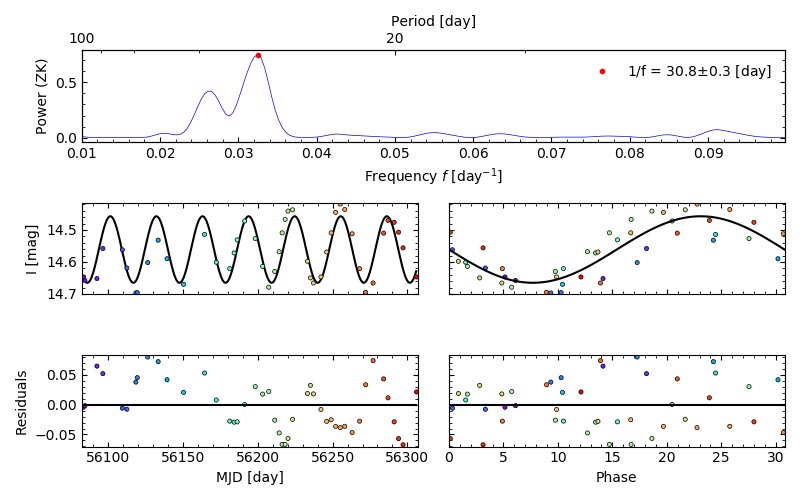}
    \plotone{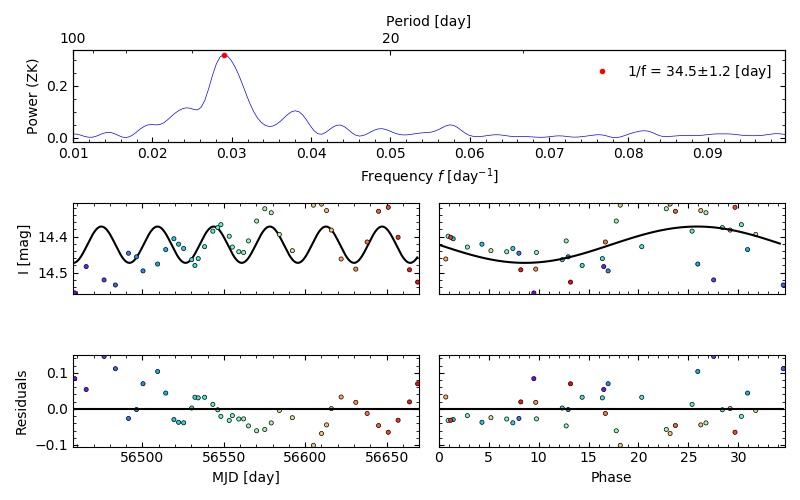}
% \vspace*{-1in}
  \caption{(Continued)}
% \label{fig:GLSfits}
\end{figure*}

\begin{figure*}[h]
\figurenum{13}
\vspace*{-0.3in}
      \plotone{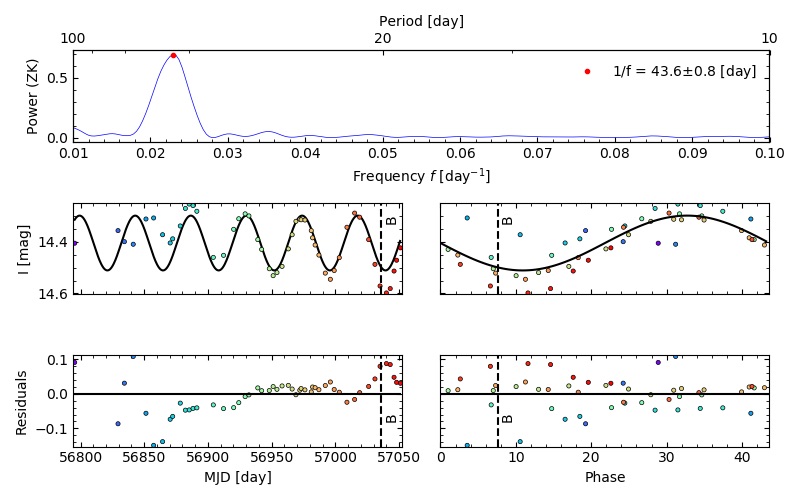}
     \plotone{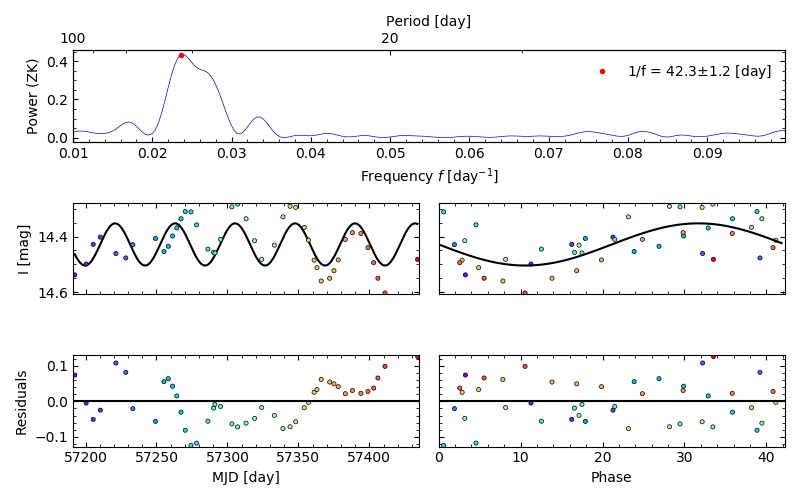}
% \vspace*{-1in}
  \caption{(Continued)}
% \label{fig:GLSfits}
\end{figure*}

\listofchanges

\end{document}